\documentclass[preprint2]{aastex63}
\usepackage{graphicx}
\usepackage{longtable}
\usepackage{natbib,twoopt}
\usepackage{url}
\usepackage[normalem]{ulem}
\def \be {\begin{equation}}
\def \en {\end{equation}}

\def \mta{}

\def \ma {}

\def\t0 {$T_0$ }     % {$10:11:58.599$}

\def \chfp{CHIME/FRB }
\def \chf{CHIME/FRB}
\def \sgrbp{SGR 1935+2154 }
\def \sgrb{SGR 1935+2154}

\def \fv {}
\def \fvv {}
\def \fvr {}
\def \cas {}

\def \cc {}

\def \chf{CHIME/FRB }
\def \chfp{CHIME/FRB}
\def \sgr{SGR 1806-20 }
\def \sgrp{SGR 1806-20}
\def \ffrbnp{180916.J0158+65}

\def \ffrbp{FRB 180916.J0158+65}
\def\ffrb{FRB 180916.J0158+65 }
\def \ffrb2p{FRB 181030.J1054+73}
\def \ffrbnpB{181030.J1054+73}
\def \ffrb2{FRB 181030.J1054+73 }

\received{December 22, 2020}
\revised{April 29, 2021}
\accepted{April 29, 2021}
\submitjournal{ApJ}
\shorttitle{\small AGILE observations of FRB Sources}
\shortauthors{\small Verrecchia et al.}

\begin{document}

\title{AGILE Observations of Fast Radio Bursts}

\author[0000-0003-3455-5082]{F. Verrecchia}
\email{francesco.verrecchia@inaf.it}
\affiliation{\scriptsize SSDC/ASI, via del Politecnico snc, I-00133 Roma (RM), Italy}
\affiliation{\scriptsize INAF/OAR, via Frascati 33, I-00078 Monte Porzio Catone (RM), Italy}

\author[0000-0001-8100-0579]{C. Casentini}
\email{claudio.casentini@inaf.it}
\affiliation{\scriptsize INAF/IAPS, via del Fosso del Cavaliere 100, I-00133 Roma (RM), Italy}
\affiliation{\scriptsize INFN Sezione di Roma 2, via della Ricerca Scientifica 1, I-00133 Roma (RM), Italy}

\author[0000-0003-2893-1459]{M. Tavani}
\email{marco.tavani@inaf.it}
\affiliation{\scriptsize INAF/IAPS, via del Fosso del Cavaliere 100, I-00133 Roma (RM), Italy}
\affiliation{\scriptsize Universit\`a degli Studi di Roma Tor Vergata, via della Ricerca Scientifica 1, I-00133 Roma (RM), Italy}

\author[0000-0002-7253-9721]{A. Ursi}
%\email{alessandro.ursi@inaf.it}
\affiliation{\scriptsize INAF/IAPS, via del Fosso del Cavaliere 100, I-00133 Roma (RM), Italy}

\author[0000-0003-3259-7801]{S. Mereghetti}
%\email{sandro.mereghetti@inaf.it}
\affiliation{\scriptsize INAF/IASF, via E. Bassini 15, I-20133 Milano (MI), Italy}

\author[0000-0001-7397-8091]{M. Pilia}
\affiliation{\scriptsize INAF/OAC, via della Scienza 5, I-09047 Selargius (CA), Italy}
%\email{maura.pilia@inaf.it}

\author[0000-0001-8877-3996]{M. Cardillo}
%\email{martina.cardillo@inaf.it}
\affiliation{\scriptsize INAF/IAPS, via del Fosso del Cavaliere 100, I-00133 Roma (RM), Italy}

\author{A.~Addis}
%\email{antonio.addis@inaf.it}
\affiliation{\scriptsize INAF/OAS, via Gobetti 101, I-40129 Bologna (BO), Italy}

\author{G. Barbiellini}
%\email{guido.barbiellini@ts.infn.it}
\affiliation{\scriptsize Dipartimento di Fisica, Universit\`a di Trieste and INFN, via Valerio 2, I-34127 Trieste (TS), Italy}

\author{L.~Baroncelli}
%\email{leonardo.baroncelli@inaf.it}
\affiliation{\scriptsize INAF/OAS, via Gobetti 101, I-40129 Bologna (BO), Italy}
\affiliation{\scriptsize Dip. di Fisica e Astronomia, Universit\`a di Bologna, Viale Berti Pichat 6/2, 40127, Bologna, Italy}

\author[0000-0001-6347-0649]{A. Bulgarelli}
%\email{andrea.bulgarelli@inaf.it}
\affiliation{\scriptsize INAF/OAS, via Gobetti 101, I-40129 Bologna (BO), Italy}

\author[0000-0001-6877-6882]{P.W. Cattaneo}
%\email{paolo.cattaneo@pv.infn.it}
\affiliation{\scriptsize INFN Sezione di Pavia, via U. Bassi 6, I-27100 Pavia (PV), Italy}

\author[0000-0001-6425-5692]{A. Chen}
%\email{Andrew.Chen@wits.ac.za}
\affiliation{\scriptsize School of Physics, Wits University, Johannesburg, South Africa}

\author[0000-0003-4925-8523]{E. Costa}
%\email{enrico.costa@iaps.inaf.it}
\affiliation{\scriptsize INAF/IAPS, via del Fosso del Cavaliere 100, I-00133 Roma (RM), Italy}

\author[0000-0002-3013-6334]{E. Del Monte}
%\email{ettore.delmonte@inaf.it}
\affiliation{\scriptsize INAF/IAPS, via del Fosso del Cavaliere 100, I-00133 Roma (RM), Italy}

\author[0000-0002-9894-7491]{A.~Di Piano}
%\email{ambra.dipiano@inaf.it}
\affiliation{\scriptsize INAF/OAS, via Gobetti 101, I-40129 Bologna (BO), Italy}

\author{A. Ferrari}
%\email{ferrari@ph.unito.it}
\affiliation{\scriptsize CIFS, c/o Physics Department, University of Turin, via P. Giuria 1, I-10125,  Torino, Italy}

\author[0000-0002-6082-5384]{V.~Fioretti}
%\email{valentina.fioretti@inaf.it}
\affiliation{\scriptsize INAF/OAS, via Gobetti 101, I-40129 Bologna (BO), Italy}

\author[0000-0003-2501-2270]{F. Longo}
%\email{franzlongo1969@gmail.com}
\affiliation{\scriptsize Dipartimento di Fisica, Universit\`a di Trieste and INFN, via Valerio 2, I-34127 Trieste (TS), Italy}

\author[0000-0002-6311-764X]{F. Lucarelli}
%\email{fabrizio.lucarelli@ssdc.asi.it}
\affiliation{\scriptsize SSDC/ASI, via del Politecnico snc, I-00133 Roma (RM), Italy}
\affiliation{\scriptsize INAF/OAR, via Frascati 33, I-00078 Monte Porzio Catone (RM), Italy}

\author[0000-0002-4535-5329]{N. Parmiggiani}
%\email{nicolo.parmiggiani@inaf.it}
\affiliation{\scriptsize INAF/OAS, via Gobetti 101, I-40129 Bologna (BO), Italy}
\affiliation{Universit\'a degli Studi di Modena e Reggio Emilia, DIEF - Via Pietro Vivarelli 10, 41125 Modena, Italy}

\author[0000-0002-9332-5319]{G. Piano}
%\email{giovanni.piano@inaf.it}
\affiliation{\scriptsize INAF/IAPS, via del Fosso del Cavaliere 100, I-00133 Roma (RM), Italy}

\author[0000-0001-6661-9779]{C. Pittori}
%\email{carlotta.pittori@ssdc.asi.it}
\affiliation{\scriptsize SSDC/ASI, via del Politecnico snc, I-00133 Roma (RM), Italy}
\affiliation{\scriptsize INAF/OAR, via Frascati 33, I-00078 Monte Porzio Catone (RM), Italy}

\author[0000-0001-9702-7645]{A. Rappoldi}
%\email{paolo.cattaneo@pv.infn.it}
\affiliation{\scriptsize INFN Sezione di Pavia, via U. Bassi 6, I-27100 Pavia (PV), Italy}

\author[0000-0003-1163-1396]{S. Vercellone}
%\email{stefano.vercellone@inaf.it}
\affiliation{\scriptsize INAF/OAB, via E. Bianchi 46, I-23807 Merate (LC), Italy}

\begin{abstract}
%\abstract{}
{\ma We report on a systematic search for hard X-ray and $\gamma$-ray emission in coincidence with fast radio bursts (FRBs) observed by the AGILE satellite}.
{\ma We used 13 yr of AGILE archival data searching for time coincidences between exposed FRBs and events detectable by the MCAL (0.4\,-\,100 MeV) and GRID (50 MeV\,-\,30 GeV) detectors at timescales ranging from milliseconds to days/weeks. The current AGILE sky coverage allow{\fvv ed} us to extend the search for high-energy emission {\fvv preceding} and {\fvv following} the FRB occurrence.}
{\ma We considered all FRBs sources currently included in catalogs, and identified} a sub{\fvv sample} (1{\fvv 5} events) for which a {\fvv good} AGILE exposure either with MCAL or GRID was obtained. In this paper we focus on nonrepeating FRBs{\fvr , compared to a few nearby repeating sources}.
{\ma
We did not detect significant MeV or GeV emission from any event.}
{\ma Our hard X-ray upper limits (ULs) in the MeV energy range were obtained for timescales from submillisecond to seconds, and in the GeV range from minutes to weeks around event times.}
We focus on a subset of {\fvv five} non-repeating and two repeating FRB sources whose distances are most likely smaller than that of
%{\mta FRB}
{\ffrbnp} (150 Mpc). For these sources, our MeV ULs translate into ULs on the isotropically emitted energy {\fvv of about} $3\,\times\,10^{46} \, \rm erg$,
comparable to that observed {\fvv in} the 2004 giant {\fvv flare} from the Galactic magnetar \sgrp. On average, these nearby FRBs emit radio pulses of energies significantly larger than the recently detected \sgrbp and are not yet associated with intense MeV flaring.
%}

\end{abstract}

%\maketitle  %AA

    %%%%%%%%%%%%%%%%%%%%%%%%%%%%%%%%%%%%%%%%%%%%%%%%%%%%%%%%%%%%%%%%%%%%
    \section{Introduction}
    %%%%%%%%%%%%%%%%%%%%%%%%%%%%%%%%%%%%%%%%%%%%%%%%%%%%%%%%%%%%%%%%%%%%

%
%This event is the fifth of a set of confirmed GW events detected by
%
Fast radio bursts (FRBs) are {\ma transient events of unknown origin} consisting of bright millisecond radio pulses {\fvr at $\sim 1\, \rm GHz$} having usually large dispersion measures (DMs; the free electron density along the line of sight at a certain distance) in excess of Galactic values (\citealt[][]{2007Science.218.777,2019MRAS487..491K,2019A&ARv27..4P}{\fvv , and }\citealt[][{\fvv hereafter CC19}]{2019ARA&A119..161101}). The sources known so far, 110 as of 2020 May 1 (see FRBCAT\footnote{http://www.frbcat.org/} \citealt[][]{2016PASA33..e45}, {\fvv and the \chf telescope public on-line database\footnote{https://www.chime-frb.ca}}), have an isotropic sky distribution and for most of them a single radio burst %has been
{\mta was} detected. A smaller, but increasing number of FRBs show repeating pulses{ \fvv \citep[][]{2016Natur.531.202S},} which {\ma usually} occur erratically in time, {\fvv and two of them show a periodic pattern in their activity cycle,} {\mta as in} the case of {\ffrbp} \cite[][{\mta hereafter} C20]{2020ApJLCHIME}. No clear evidence for physically different populations {\ma distinguishing repeating and nonrepeating sources} has been obtained so far.
% Due to the uncertain distance of most FRBs, their distances
{\ma The FRB distances can be deduced from their DM measures, which are typically DM\,$\geq$\,500\,$\rm pc \, cm^{-{\fvv 3}}${\fvv , which, taking into account Galactic and extragalactic contributions, point to an extragalactic origin.}}
%which is the sum of several contributions from our Galaxy and from the host galaxy and } are usually extrapolated from the DM measure, which includes however the uncertain host galaxy component.
The short duration of {\ma FRBs (typically milliseconds or less)} {\ma favor{\fvv s} models involving compact objects such as strongly magnetized neutron stars (magnetars) and massive black holes}
%favorite models involving stellar mass compact objects, such as pulsars or magnetars
\citep[][]{2019PhR821..1P}.

{\fvv Since{ \fvr 2019} {\mta the} \chf collaboration} has reported the discovery of
% an additional repeating FRB, FRB180814.J0453
{\ma several repeating FRBs (hereafter R-FRBs;} \citealt[][{\fvv hereafter C19}]{2019ApJLCHIME,2019ApJLCHIMEb,2019ApJLCHIMEc}; \citealt[][]{2020ApJL.891..L6F}),
%\citep[][]{2020ApJL.891..L6F}
with DM ranging from 103.5 to 1378\,$\rm pc\,cm^{-3}$. The characteristics of this sample of R-FRBs are particularly interesting for counterpart searches, as we discussed in \citealt[][]{2020Casentini}{\ma {\fvv ,} focused on} two nearby sources, {\ffrbnp} (Source 1, {\ma following the nomenclature of C19}) and {\ffrbnpB} (Source 2). These two sources have low DM values in excess of their respective Galactic disk and halo contributions (see next section) {\ma and are therefore expected to be relatively close to Earth (as confirmed by the redshift-determined distance of {\fvv Source 1} equal to 149 Mpc; \citealt[][hereafter M20]{2020Natur577...190}). As we will see, this distance determination for a prominent FRB sets the scale for our discussion on the implications of our search for high-energy emission from FRBs}.
{\ma Source 1} has been recently reported to have periodic bursting activity (C20), with most radio bursts
%{\ma being}
emitted within time intervals recurring periodically {\ma every 16.{\fvr 3} days} \citep[][]{2021ApJLPleunis}. A multifrequency follow-up campaign {\ma has been} organized to search for high-energy and optical emission during {\ma active 
\pagebreak
%\newpage
%%% %%\pagebreak
%\input{AGFRB_tablev3.tex}
%\input{AGFRB_tablev4.tex}
%\input{AGFRB_tablev5.2.tex}
%\input{AGFRB_tablev6.tex}
%\input{AGFRB_tablev7.tex}
%\input{AGFRB_tablev8.tex}
%\input{AGFRB_tablev9d4.tex}
%\input{AGFRB_tablev9d5.tex}
%\onecolumn
%\newpage
%\pagebreak

% \begin{landscape}
%\begin{sidewaystable}
%\tiny
%\scriptsize
%\fontsize{6}{6}\selectfont
%{\setlength{\tabcolsep}{2pt}
%\rotatebox{90}{
%\longtab{1}{
%\LTcapwidth=10.2in

%\begin{longtable}{lcccccccccl}
%\begin{longtable}{lccccccccccl}
%\begin{longtable}{lccccccclllll}
%\begin{longtable}{lccccccclll}
%\begin{table*}
\begin{longrotatetable}
%\begin{rotatetable}
%\begin{deluxetable*}{lcccccccllll}
%\noindent\makebox[\textwidth]{%
%\startlongtable
\begin{deluxetable*}{lccccccclll}
\tablewidth{700pt}
\tabletypesize{\scriptsize}
%  \caption{FRB Catalog sources. See Table for a description of the columns. Events with $*$ occur before AGILE operational lifetime.}\\
  \tablecaption{FRB Catalog Sources.}
  \label{tab-1}
%  \midrule[0.6pt]
%  \toprule[1pt]
%\begin{tabular}{lcccccccccccl}

% & & & & & &  & & & & &   &     \\[3 pt]
%\multicolumn{1}{|l}{} &
%\multicolumn{1}{c}{} &
%\multicolumn{1}{c}{} &
%% added
%\multicolumn{1}{c}{} &
%\multicolumn{1}{c}{} &
%%
%\multicolumn{1}{c}{} &
%\multicolumn{1}{c}{} &
%% added
%\multicolumn{1}{c}{} &
%%
%\multicolumn{1}{c}{} &
%\multicolumn{1}{c}{} &
%% added
%\multicolumn{1}{c}{} &
%\multicolumn{1}{c}{} &
%\multicolumn{1}{c|}{Possible Counterp. }      \\[3 pt]
%%%%%%%%%%%%%%%%%%%%%%%%
%\hline
\tablehead{
\colhead{ID} &
% added
\colhead{DATE } &
%\multicolumn{1}{c}{Dec.} &
\colhead{\textit{LII}} &
\colhead{\textit{BII}} &
\colhead{DM and Error} &
%\colhead{Radio Peak} &
\colhead{Radio } &
\colhead{Within} &
\colhead{Within} &
\colhead{Within} &
% added
\colhead{{Off-axis}} &
%\colhead{{Occulted}} &
\colhead{{Repeater}} \\[3 pt]
%\multicolumn{1}{c}{Original Source} &
%\multicolumn{1}{c}{Distance} \\[3 pt]
%\multicolumn{1}{c}{Notes}
%%%%%
\colhead{{~~~}} &
% added
\colhead{{UTC}} &
\colhead{{ }} &
\colhead{{ }} &
\colhead{{  }} &
\colhead{{Fluence}} &
\colhead{{MCAL FOV}} &
\colhead{{GRID FOV}} &
\colhead{{Super-A FOV}} &
% added
\colhead{{Angle}} &
%\colhead{{by}} &
\colhead{{~~~ }} \\[3 pt]
%%%%%%%%%%%%%%%%%%%%%%%%%%%%%%%%%%%%%
\colhead{{ }} &
% added
\colhead{{~~~ }} &
\colhead{{ (deg) }} &
\colhead{{ (deg) }} &
\colhead{{ ($\rm cm^{-3} pc$) }} &
\colhead{{ (Jy ms) }} &
%\multicolumn{1}{c}{ (deg)} &
\colhead{{~~~}} &
% added
\colhead{{~~~}} &
%
%\multicolumn{1}{c}{  ($\times 10^{-8}$ ph cm$^{-2}$ s$^{-1}$) } %&
%\multicolumn{1}{c}{Flux Error} &
% added
\colhead{{~~~}} &
\colhead{{ (deg)}} &
\colhead{{~~~}}  %\\[3 pt]
%\multicolumn{1}{c}{}      \\[3 pt]%&
%\multicolumn{1}{c}{} (Mpc)   \\[3 pt]
%\multicolumn{1}{c|}{3EG Counterp.}      \\[3 pt]
%\toprule[0.6pt]
% & & & &  & & & & & &     \\[3 pt]
%\hline
% \endhead
}
%\toprule[1pt]
%\hline
%\endfoot
\startdata
% & & & & & &  & & & & &      \\%&    \\[3 pt]
%,FRB_NAME,FRB_UTC_TIME,g.l.,g.b.,UNOCCULTED,GRID F.O.V.,SA F.O.V.,OFF-AXIS,OCCULTED BY,REPEATER,ORIGINAL SOURCE
%010125$^{*}$     &  2001-01-25 00:29:15.790  &  356.64 &  -20.02 &  790\,$\pm$\,3         &  2.8200  &   -    &    -   &    -   &   0.00 &  -
%010312$^{*}$     &  2001-03-12 11:06:47.980  &  274.72 &  -33.30 &  1187\,$\pm$\,14       &  6.0750  &   -    &    -   &    -   &   0.00 &  -
%010621$^{*}$     &  2001-06-21 13:02:11.299  &   25.43 &   -4.00 &  745\,$\pm$\,10        &  2.8700  &   -    &    -   &    -   &   0.00 &  -
%010724$^{*}$     &  2001-07-24 19:50:01.690  &  300.65 &  -41.81 &  375.00                & 150.0000 &   -    &    -   &    -   &   0.00 &  -
090625            &  2009-06-25 21:53:51.379  &  226.44 &  -60.03 &  899.55\,$\pm$\,0.01   &   2.189  &  YES   &   NO   &    NO  &  94.91 &  NO    \\           %
110214            &  2011-02-14 07:14:10.353  &  290.70 &  -66.60 &  168.9\,$\pm$\,0.5     &  51.300  &  YES   &   YES  &    NO  &  63.02 &  NO    \\           %
110220            &  2011-02-20 01:55:48.096  &   50.83 &  -54.77 &  944.38\,$\pm$\,0.05   &   7.310  &  NO    &   NO   &    NO  &  93.94 &  NO    \\           %
110523            &  2011-05-23 15:06:19.700  &   56.12 &  -37.82 &  623.3\,$\pm$\,0.06    &   1.038  &  YES   &  YES   &    NO  &  44.12 &  NO    \\           %
110626            &  2011-06-26 21:33:17.477  &  355.86 &  -41.75 &  723.0\,$\pm$\,0.3     &   0.560  &  NO    &   NO   &    NO  & 128.61 &  NO    \\           %
110703            &  2011-07-03 18:59:40.607  &   81.00 &  -59.02 &  1103.6\,$\pm$\,0.7    &   1.750  &  NO    &   NO   &    NO  & 157.70 &  NO    \\           %
120127            &  2012-01-27 08:11:21.725  &   49.29 &  -66.20 &  553.3\,$\pm$\,0.3     &   0.750  &  -     &    -   &    -   &   -    &  NO    \\           %
121002            &  2012-10-02 13:09:18.436  &  308.22 &  -26.26 &  1629.18\,$\pm$\,0.02  &   2.280  &  YES   &  YES   &    NO  &  63.80 &  NO    \\           %
%121029            &  2012-10-29 16:06:26.0    &  115.15 &  -20.22 &        ---             & 108.8000 &  IDLE  &    -   &    -   &  11.60 &  NO    \\    %added %Pushchino
121102            &  2012-11-02 06:35:53.244  &  174.95 &   -0.23 &  557\,$\pm$\,2         &  1.200   &  NO    &   NO   &    NO  & 133.28 &  YES     \\           %
130626            &  2013-06-26 14:55:59.771  &    7.45 &   27.42 &  952.4\,$\pm$\,0.1     & $>\,$1.500 &  -     &   -    &    -   &   -   &  NO    \\           %
130628            &  2013-06-28 03:58:00.178  &  225.96 &   30.66 &  469.88\,$\pm$\,0.01   & $>\,$1.220 &  YES   &  YES   &    NO  &  50.26 &  NO    \\           %
130729            &  2013-07-29 09:01:51.190  &  324.79 &   54.74 &  861\,$\pm$\,2         & $>$\,3.5   &  NO    &   NO   &    NO  &  67.85 &  NO    \\           %
%131030            &  2013-10-30 16:13:15.0    &  132.93 &  -22.61 &        ---             & 127.2000 &    YES & $\sim$\,YES  &    NO  &  71.16 &  NO      \\    %added %Pushchino
131104            &  2013-11-04 18:04:11.200  &  260.55 &  -21.93 &  779\,$\pm$\,1         &   2.750  &  -     &   -    &    -   &   -    &  NO    \\           %
%140212            &  2014-02-12 10:31:14.0    &  132.93 &  -31.56 &        ---             & 101.4000 & IDLE   &   -    &    -   &  34.92 &  NO    \\    %added %Pushchino
140514            &  2014-05-14 17:14:11.060  &   50.84 &  -54.61 &  562.7\,$\pm$\,0.6     &   1.320  &  NO    &   NO   &    NO  & 144.34 &  NO    \\           %
141113            &  2014-11-13 07:42:55.220  &  191.90 &    0.36 &  400.3                 &   0.078  &  YES   &  YES   &    NO  &  61.39 &  NO    \\           %
%141216            &  2014-12-16 13:03:24.0    &  115.47 &  -20.69 &        ---             & 200.1000 &  YES   &   NO   &    NO  & 100.77 &  NO      \\    %added %Pushchino
150215            &  2015-02-15 20:41:41.714  &   24.66 &    5.28 &  1105.6\,$\pm$\,0.8    &   2.020  &  -     &   -    &    -   &   -    &  NO    \\           %
150418            &  2015-04-18 04:29:06.657  &  232.67 &   -3.23 &  776.2\,$\pm$\,0.5     &   1.760  &  NO    &   NO   &    NO  & 161.08 &  NO    \\           %
150610            &  2015-06-10 05:26:59.396  &  278.00 &   16.50 &  1593.9\,$\pm$\,0.6    &   1.300  &  YES   &   NO   &    NO  & 141.61 &  NO    \\           %
150807            &  2015-08-07 17:53:55.830  &  333.89 &  -53.60 &  266.5\,$\pm$\,0.1     &  44.800  &  NO    &   NO   &    NO  &  49.49 &  NO    \\           %
%151018*           &  2015-10-18 01:05:48.000  &  173.60 &   -2.15 &  570\,$\pm$\,5         & 3500.0000 &  YES   &   NO   &    NO  & 132.02 &  NO    \\        %  Pushchino
%151125            &  2015-11-25 15:42:36.0    &  132.84 &  -31.13 &        ---             & 907.2000 & IDLE   &   -    &    -   & 123.34 &  NO      \\    %added % Pushchino
%%151125           &  2015-11-28 15:43:56.0    &  133.08 &  -31.09 &        ---             & 907.2000 & IDLE   &   -    &    -   &  67.43 &  NO      \\    %added % Pushchino
151206            &  2015-12-06 06:17:52.778  &   32.60 &   -8.50 &  1909.8\,$\pm$\,0.6    &   0.900  &  NO    &   NO   &    NO  &  63.28 &  NO    \\           %
151230            &  2015-12-30 16:15:46.525  &  239.00 &   34.80 &  960.4\,$\pm$\,0.5     &   1.900  &  YES   &   NO   &    NO  &  90.33 &  NO    \\           %
160102            &  2016-01-02 08:28:39.374  &   18.90 &  -60.80 &  2596.1\,$\pm$\,0.3    &   1.800  &  NO    &   NO   &    NO  &  44.87 &  NO    \\           %
%160206            &  2016-02-06 10:26:50.0    &  124.85 &  -21.21 &        ---             & 413.4000 &  YES   &   NO   &    NO  & 163.04 &  NO      \\    %added % Pushchino
160317            &  2016-03-17 09:00:36.530  &  246.05 &   -0.99 &  1165\,$\pm$\,11       &  63.000  &  NO    &   NO   &    NO  &  61.41 &  NO    \\           %
160410            &  2016-04-10 08:33:39.680  &  220.36 &   27.19 &  278\,$\pm$\,3         &  28.000  &  NO    &   NO   &    NO  &  32.96 &  NO    \\           %
160608            &  2016-06-08 03:53:01.088  &  254.11 &   -9.54 &  682\,$\pm$\,7         &  38.700  &  NO    &   NO   &    NO  & 165.22 &  NO    \\           %
%160920*           &  2016-09-20 03:05:43      &  167.80 &    4.79 &  1767\,$\pm$\,4        & 1100.0000 &  NO    &   NO   &    NO  &  90.26 &  NO    \\           % Pushchino
%161202            &  2016-12-02 13:24:54.0    &  109.40 &  -20.29 &        ---             & 234.9000 &  YES   &   NO   &    NO  & 154.98 &  NO     \\    %added % Pushchino
170107            &  2017-01-07 20:05:08.139  &  266.08 &   51.45 &  609.5\,$\pm$\,0.5     &  58.000  &  YES   &  YES   &    YES &  29.45 &  NO    \\           %
170416            &  2017-04-16 23:11:12.799  &  337.62 &  -50.05 &  523.2\,$\pm$\,0.2     &  97.000  &  YES   &   NO   &    NO  & 133.52 &  NO    \\           %
170428            &  2017-04-28 18:02:34.700  &  326.75 &  -41.85 &  991.7\,$\pm$\,0.9     &  34.000  &  YES   &   NO   &    NO  & 122.82 &  NO    \\           %
%170606*           &  2017-06-06 10:03:27.000  &  167.80 &    4.79 &  247\,$\pm$\,5         & 1782.0000 &  YES   &   NO   &    NO  &  75.82 &  NO     \\           % Pushchino
170707            &  2017-07-07 06:17:34.354  &  275.02 &  -52.39 &  235.2\,$\pm$\,0.6     &  52.000  &  YES   &  YES   &    NO  &  50.66 &  NO    \\           %
170712            &  2017-07-12 13:22:17.394  &  327.41 &  -49.24 &  312.79\,$\pm$\,0.07   &  53.000  &  YES   &   NO   &    NO  & 104.46 &  NO    \\           %
170827            &  2017-08-27 16:20:18.000  &  303.20 &  -51.70 &  176.80\,$\pm$\,0.04   &  19.870  &  NO    &   NO   &    NO  &  81.31 &  NO    \\           %
170906            &  2017-09-06 13:06:56.488  &   33.85 &  -50.20 &  390.3\,$\pm$\,0.4     &  74.000  &  YES   &   NO   &    NO  &  73.37 &  NO    \\           %
170922            &  2017-09-22 11:23:33.400  &   45.10 &  -38.70 &  1111\,$\pm$\,1        & 177.000  &  YES   &   NO   &    NO  & 128.74 &  NO    \\           %
171003            &  2017-10-03 04:07:23.781  &  294.90 &   48.41 &  463.2\,$\pm$\,1.2     &  81.000  &  YES   &   NO   &    NO  &  95.98 &  NO    \\           %
171004            &  2017-10-04 03:23:39.250  &  282.69 &   48.84 &  304.0\,$\pm$\,0.3     &  44.000  &  YES   &   NO   &    NO  &  79.72 &  NO    \\           %
171019            &  2017-10-19 13:26:40.097  &   52.51 &  -49.24 &  460.8\,$\pm$\,1.1     & 219.000  &  YES   &   NO   &    NO  & 134.78 &  NO    \\           %
171020            &  2017-10-20 10:27:58.598  &   36.17 &  -53.49 &  114.1\,$\pm$\,0.2     & 200.000  &  NO    &   NO   &    NO  &  80.64 &  NO    \\           %
171116            &  2017-11-16 14:59:33.305  &  206.37 &  -51.93 &  618.5\,$\pm$\,0.5     &  63.000  &  YES   &   NO   &    NO  &  95.06 &  NO    \\           %
171209            &  2017-12-09 20:34:23.500  &  332.20 &    6.24 &  1457.40\,$\pm$\,0.03  &   3.700  &  YES   &   NO   &    NO  & 105.10 &  NO    \\           %
171213            &  2017-12-13 14:22:40.467  &  198.85 &  -47.48 &  158.6\,$\pm$\,0.2     & 133.000  &  YES   &   NO   &    NO  & 126.69 &  NO    \\           %
171216            &  2017-12-16 17:59:10.822  &  271.19 &  -49.29 &  203.1\,$\pm$\,0.5     &  40.000  &  NO    &   NO   &    NO  &  30.35 &  NO    \\           %
180110            &  2018-01-10 07:34:34.959  &    9.18 &  -51.32 &  715.7\,$\pm$\,0.2     & 420.000  &  YES   &   NO   &    NO  & 121.76 &  NO    \\           %
180119            &  2018-01-19 12:24:40.747  &  199.58 &  -50.41 &  402.7\,$\pm$\,0.7     & 110.000  &  YES   &   NO   &    NO  &  98.08 &  NO    \\           %
180128.0          &  2018-01-28 00:59:38.617  &  330.05 &   52.72 &  441.4\,$\pm$\,0.2     &  51.000  &  YES   &   NO   &    NO  & 121.32 &  NO    \\           %
180128.2          &  2018-01-28 04:53:26.796  &  329.86 &  -48.32 &  495.9\,$\pm$\,0.7     &  66.000  &  YES   &   YES  &    NO  &  39.87 &  NO    \\           %
180130            &  2018-01-30 04:55:29.993  &    4.21 &  -51.07 &  343.5\,$\pm$\,0.4     &  95.000  &  -     &   -    &    -   &    -   &  NO    \\           %
180131            &  2018-01-31 05:45:04.320  &    0.93 &  -50.47 &  657.7\,$\pm$\,0.5     & 100.000  &  YES   &   NO   &    NO  & 116.37 &  NO    \\           %
180212            &  2018-02-12 23:45:04.399  &  341.52 &   52.45 &  167.5\,$\pm$\,0.5     &  96.000  &  IDLE  &   NO   &    NO  &  54.69 &  NO    \\           %
180309            &  2018-03-09 02:49:32.990  &   10.90 &  -45.40 &  263.42\,$\pm$\,0.01   &  13.120  &  YES   &   YES  &    NO  &  54.96 &  NO    \\           %
180311            &  2018-03-11 04:11:54.800  &  337.30 &  -43.70 &  1570.9\,$\pm$\,0.5    &   2.100  &  IDLE  &   -    &     -  &  45.81 &  NO    \\           %
180315            &  2018-03-15 05:05:30.985  &   13.20 &  -20.90 &  479.0\,$\pm$\,0.4     &  56.000  &   NO   &   NO   &    NO  & 148.08 &  NO    \\           %
%180321            &  2018-03-21 07:05:54.0    &  119.27 &  -20.71 &        ---             & 901.8000 &  NO    &   NO   &    NO  & 106.30 &  NO    \\    %added % Pushchino
180324            &  2018-03-24 09:31:46.706  &  245.20 &  -20.50 &  431.0\,$\pm$\,0.4     &  71.000  &  YES   &   NO   &    NO  & 126.04 &  NO    \\           %
180417            &  2018-04-17 13:18:31.000  &  276.00 &   75.60 &  474.8                 &  55.000  &  YES   &   NO   &    NO  & 105.53 &  NO    \\           %
180430            &  2018-04-30 09:59:58.700  &  221.76 &   -4.61 &  264.1\,$\pm$\,0.5     & 177.000  &  YES   &   NO   &    NO  & 168.04 &  NO    \\           %
180515            &  2018-05-15 21:57:26.485  &  349.50 &  -64.90 &  355.2\,$\pm$\,0.5     &  46.000  &  YES   &   YES  &    NO  &  60.12 &  NO    \\           %
180525            &  2018-05-25 15:19:06.515  &  349.00 &   50.70 &  388.1\,$\pm$\,0.3     & 300.000  &  NO    &   NO   &    NO  &  60.90 &  NO    \\           %
180528            &  2018-05-28 04:24:00.900  &  258.80 &  -22.35 &  899.3\,$\pm$\,0.6     &  32.000  &  YES   &   NO   &    NO  & 111.25 &  NO    \\           %
180714            &  2018-07-14 10:00:08.700  &   14.80 &    8.72 &  1467.92\,$\pm$\,0.3   &   1.850  &  NO    &   NO   &    NO  &  72.72 &  NO    \\           %
180725.J0613+67   &  2018-07-25 17:59:32.813  &  147.00 &   21.00 &  715.98\,$\pm$\,0.2    &  12.000  &  YES   &   NO   &    NO  & 114.43 &  NO    \\           %
180727.J1311+26   &  2018-07-27 00:52:04.474  &   25.00 &   85.00 &  642.07\,$\pm$\,0.03   &  14.000  &  YES   &   NO   &    NO  & 152.33 &  NO    \\           %
180729.J0558+56   &  2018-07-29 00:48:19.238  &  115.00 &   61.00 &  317.37\,$\pm$\,0.01   &  34.000  &  YES   &   YES  &    NO  &  61.89 &  NO    \\           %
180729.J1316+55   &  2018-07-29 17:28:18.258  &  156.00 &   15.00 &  109.610\,$\pm$\,0.002 &   9.000  &  YES   &   NO   &    NO  & 100.32 &  NO    \\           %
180730.J0353+87   &  2018-07-30 03:37:25.937  &  125.00 &   25.00 &  849.047\,$\pm$\,0.002 &  50.000  &  YES   &   NO   &    NO  & 155.51 &  NO    \\           %
180801.J2130+72   &  2018-08-01 08:47:14.793  &  109.00 &   15.00 &  656.20\,$\pm$\,0.03   &  28.000  &  NO    &   NO   &    NO  &  42.06 &  NO    \\           %
180806.J1515+75   &  2018-08-06 14:13:03.107  &  112.00 &   38.00 &  739.98\,$\pm$\,0.03   &  24.000  &  YES   &   YES  &    NO  &  62.78 &  NO    \\           %
180810.J0646+34   &  2018-08-10 17:28:54.614  &  180.00 &   14.00 &  414.95\,$\pm$\,0.02   &  11.000  &  NO    &   NO   &    NO  &  64.62 &  NO    \\           %
180810.J1159+83   &  2018-08-10 22:40:42.493  &  125.00 &   34.00 &  169.134\,$\pm$\,0.002 &  17.000  &  YES   &   NO   &    NO  & 117.77 &  NO    \\           %
180812.J0112+80   &  2018-08-12 11:45:32.872  &  123.00 &   18.00 &  802.57\,$\pm$\,0.04   &  18.000  &  YES   &   NO   &    NO  & 153.02 &  NO    \\           %
180814.J1554+74   &  2018-08-14 14:20:14.440  &  108.00 &   37.00 &  238.32\,$\pm$\,0.01   &  25.000  &  YES   &   NO   &    NO  & 165.94 &  NO    \\           %
180814.J0422+73   &  2018-08-14 14:49:48.022  &  136.00 &   16.00 &  189.38\,$\pm$\,0.09   &  21.000  &  YES   &   NO   &    NO  & 116.42 &  YES   \\           %
180817.J1533+42   &  2018-08-17 01:49:20.202  &   68.00 &   54.00 & 1006.840\,$\pm$\,0.002 &  26.000  &  NO    &   NO   &    NO  &  20.36 &  NO    \\           %
180908.J1232+74   &  2018-09-18 21:13:01.00   &  124.7  &	42.9  &  195.7\,$\pm$\,0.9     &   2.700  &  YES   &   NO   &    NO  &  81.89 &  YES   \\
180916.J0158+65   &  2018-09-16 10:15:19.802  &  129.70 &    3.70 &  349.2\,$\pm$\,0.4     &   2.30   &  NO    &   NO   &    NO  & 155.08 &  YES   \\           %
180924            &  2018-09-24 16:23:12.626  &    0.74 &  -49.41 &  361.42\,$\pm$\,0.06   &  16.000  &  YES   &   NO   &    NO  & 139.88 &  NO    \\           %
181016            &  2018-10-16 04:16:56.3    &  345.51 &   22.67 &  1982.8\,$\pm$\,2.8    &  87.000  &  YES   &   NO   &    NO  &  86.53 &  NO    \\           %
181017            &  2018-10-17 10:24:37.4    &   50.50 &  -47.00 &  239.97\,$\pm$\,0.03   &  31.000  &  YES   &   NO   &    NO  &  73.46 &  NO    \\           %
181017.J1705+68   &  2018-10-17 23:26:11.860  &   99.20 &   34.80 &  1281.9\,$\pm$\,0.4    &   1.000  &  NO    &   NO   &    NO  &   3.95 &  YES   \\	% border      %
%181017.J18+81     &  2018-10-17 16:17:19.000  &  113.3  &   27.8  &  301.6\,$\pm$\,0.3     &   3.000  &   -    &   -    &    -   &    -   &  YES    \\	      %??? 2020MNRAS 498, 3927
181030.J1054+73   &  2018-10-30 04:13:13.025  &  133.40 &   40.90 &  103.5\,$\pm$\,0.7     &   7.300  &  YES   &   NO   &    NO  &  91.44 &  YES    \\           %
181112            &  2018-11-12 17:31:15.483  &  342.60 &  -47.70 &  589.27\,$\pm$\,0.03   &  26.000  &  YES   &   YES  &    NO  &  42.93 &  NO     \\	         %
181119.J12+65     &  2018-11-19 16:49:03.191  &  124.50 &   52.00 &  364.2\,$\pm$\,1.0     &   1.800  &  NO    &   NO   &    NO  &  53.99 &  YES    \\           %
181128.J0456+63   &  2018-11-28 08:27:41.740  &  146.60 &   12.40 &  450.2\,$\pm$\,0.3     &   4.400  &   -    &   -    &    -   &    -   &  YES    \\           %
181228            &  2018-12-28 13:48:50.1    &  253.39 &  -26.06 &  354.2\,$\pm$\,0.9     &  24.000  &  YES   &   NO   &    NO  & 159.68 &  NO     \\           %
190116.J1249+27   &  2019-01-16 13:07:33.833  &  210.50 &   89.50 &  444.0\,$\pm$\,0.6     &   0.800  &  NO    &   NO   &    NO  &  48.06 &  YES    \\	         %
%190110.J1353+48  &  2019-03-03  11:05:14     &   97.5  &   65.72 &  21.8\,$\pm$\,0.5      & --       &  NO    &   NO   &    NO  &  48.06 &  YES    \\	
%190209.J0937+77   &  2019-02-09 08:20:20.977  &  134.20 &   34.80 &  424.6\,$\pm$\,0.6     &   2.00   &  YES   &   NO   &    NO  &  77.81 &  YES    \\         %
190222.J2052+69	  &  2019-02-22 18:46:01.367  &  104.90 &   15.90 &  460.6\,$\pm$\,0.1     & --       &  YES   &   YES  &    NO  &  40.38 &  YES    \\         %
190523            &  2019-05-23 06:05:55.815  &  117.03 &   44.00 &  760.8\,$\pm$\,0.6     & --       &  YES   &   NO   &    NO  & 111.84 &  NO     \\         %
%\pagebreak
190102           & 2019-01-02 05:38:43.49184  & 312.654 & -33.493 & 363.6\,$\pm$\,0.3      &  14.000  &  YES   &   NO   &    NO  &  93.41 &  NO     \\  % 0.291 1.5−1.5+3.4
190608           & 2019-06-08 22:48:12.88391  &  53.209 & -48.530 & 338.7\,$\pm$\,0.5      &  26.000  &  NO    &   NO   &    NO  & 169.31 &  NO     \\  % 0.1178 7.0 ± 1.3
190611           & 2019-06-11 05:45:43.29937  & 312.935 & -33.282 & 321.4\,$\pm$\,0.2      &  10.000  &  YES   &   YES  &    NO  &  42.93 &  NO     \\  % 0.378 17.2 ± 4.9
190711           & 2019-07-11 01:53:41.09338  & 310.908 & -33.902 & 593.1\,$\pm$\,0.4      &  34.000  &  YES   &   NO   &    NO  & 142.35 &  NO     \\  % 0.522 1.5−1.5+3.6
%190102 & 2019-01-02 05:38:43.49184 & 363.6\,$\pm$\,0.3 &  21:29:39.76 ± 0.06 ± 0.16 −79:28:32.5 ± 0.2 ± 0.5   & 14\,$\pm$\,1 &  &  &  &  &  &  \\  % 0.291 1.5−1.5+3.4
%190608 & 2019-06-08 22:48:12.88391 & 338.7\,$\pm$\,0.5 &  22:16:04.75 ± 0.02 ± 0.02 −07:53:53.6 ± 0.3 ± 0.3   & 26\,$\pm$\,4 &  &  &  &  &  &  \\  % 0.1178 7.0 ± 1.3
%190611 & 2019-06-11 05:45:43.29937 & 321.4\,$\pm$\,0.2 &  21:22:58.91 ± 0.11 ± 0.23 −79:23:51.3 ± 0.3 ± 0.6   & 10\,$\pm$\,2 &  &  &  &  &  &  \\  % 0.378 17.2 ± 4.9
%190711 & 2019-07-11 01:53:41.09338 & 593.1\,$\pm$\,0.4 &  21:57:40.68 ± 0.051 ± 0.15 −80:21:28.8 ± 0.08 ± 0.3 & 34\,$\pm$\,3 &  &  &  &  &  &  \\  % 0.522 1.5−1.5+3.6
% & & & & & &  & & & & &       \\%&    \\[3 pt]
\enddata
%\hline
%\end{tabular}
%\vspace{+0.2cm}
%\footnotetext{{{\fvv \noindent{\bf Note:} No radio fluence available in FRBCAT for 190222.J2052+69 and 190523, and no DM error for 141113 and 180417.}}}
\tablenotemark{}
\tablenotetext{}{{{\fvv \noindent{\bf Note:} No radio fluence available in FRBCAT for 190222.J2052+69 and 190523, and no DM error for 141113 and 180417.}}}
\label{tab:tab-1}
%\end{longtable}
\end{deluxetable*}
%\vspace{+0.2cm}
%\multicolumn{11}{
%\tablewidth{700pt}
%{\scriptsize{{\fvv \noindent{\bf Note:}\\ \hspace{+0.5cm}~No radio fluence available in FRBCAT for 190222.J2052+69 and 190523, and no DM error for 141113 and 180417.}}}
%{\scriptsize{{\fvv \noindent{\bf Note:} \hspace{+0.5cm}~No radio fluence available in FRBCAT for 190222.J2052+69 and 190523, and no DM error for 141113 and 180417.}}}
%\footnotesize{\centering{{\fvv \noindent{\bf Note:} No radio fluence available in FRBCAT for 190222.J2052+69 and 190523, and no DM error for 141113 and 180417.}}}\\
%}
%\multicolumn{11}{|ccccccccccc|}{{\scriptsize{{\fvv \noindent{\bf Note:}\\ \hspace{+0.5cm}~No radio fluence available in FRBCAT for 190222.J2052+69 and 190523, and no DM error for 141113 and 180417.}}} \\
%\flushleft
%\vspace{-0.1cm}
%\noindent {\bf \scriptsize Notes.\\}
%\begin{minipage}{200mm}
%{\scriptsize{{\fvv ~No radio fluence available in FRBCAT for 190222.J2052+69 and 190523, and no $DM$ error for 141113 and 180417.}}}\\
%\end{minipage}
%\end{rotatetable}
\end{longrotatetable}
%\end{table*}

%}
%}%
%\end{sidewaystable}

%\end{landscape}
%\end{rotatetable}

% reimposto il font normale
%\normalsize

% torno in modalità twocolumn
%\twocolumn

%\input{AGFRB_tablev9d6nor.tex}
%\label{tab:tab-1}
\newpage
%\KOMAoptions{paper=portrait,pagesize}
%\recalctypearea
% reimposto il font normale
%\normalsize
% torno in modalità twocolumn
%\twocolumn
phases in the period} 2020 February\,--\,September.
{\ma No X-ray emission has been detected simultaneously with radio burst activity and interesting constraints have been determined}
\citep[][]{2020Pilia,Tavani2020,2020ApJLScholz}.
Recently, {\fvr some} high-energy counterpart searches for FRBs have been {\fvr published \citep[see][]{2019ApJ879...40,2019A&A...631A..62M,2020A&A...637A..69G,2020Natur577...190,2021Univ....7...76N}, based on the updated sample of sources in the FRBCAT catalog and reporting searches in timescales from 0.1 to 100\,s in the X--ray/soft $\gamma$-ray bands. No significant detection was found, while average isotropic luminosity upper limits (ULs) have been set from Insight/HXMT data (40\,keV\,--\,3 MeV), to $\sim \,10^{49}\,$--$\,10^{51}\,\rm erg\,s^{-1}$ for the most part of FRBs and $\sim \,10^{47}\,\rm erg\,s^{-1}$ for Source 1 at 1 s integration, and 2 orders of magnitude higher at 0.1 s, while from Fermi and Swift/BAT data (8\,keV\,--\,100 GeV) lower limits to the radio-to-X--ray flux ratio at $\sim \,10^{5}\,$--$\,10^{7}\,\rm Jy\, ms\,cm^{2}\,erg^{-1}$ have been set for the nonrepeating sources}.
{\ma In this paper, we report on a systematic search in AGILE archival data for MeV and GeV emission in coincidence with FRB radio detections.}
% ...][]{2016PASA33..e45}
{\ma We searched for simultaneous or temporally close high-energy emission} for {\fvv a} sample of {\fvv 89} FRB sources selected from FRBCAT {\fvv and} from the \chf telescope public online database{\fvv ,} {\ma with the
inclusion of recent FRB detections} \citep[][]{2020Natur581...391}.{ \ma Most of these FRBs are nonrepeating.}
%focusing in particular on the nonrepeating sources. A comparison with the two R-FRBs Source 1 and 2 will be also reported.

In this paper{\fvv ,} %we present the main results of the analysis of \agile data.
{\fvv Section~2 presents a discussion %decomposition of the
{\mta of the different components making the observed DM of FRBs}. Section~3 {\ma briefly presents the characteristics of the AGILE instrument}}.
%the DM components and the application to the source discussed in the following sections, while
In Section~4 we presents the FRB list and the details of {\ma our search} analysis procedure.
We discuss our measurements and their implications with respect to the radio fluences and DM distributions.{ \mta In} Section~4.1 {\fvv and in Section~4.2 we report {\mta our} main conclusions}.
%\vspace{-1.0cm}
{\fvr
\begin{figure*}[t]
%\vspace{-1cm}
   \centerline{
\includegraphics[width=0.90\textwidth, angle = 0]{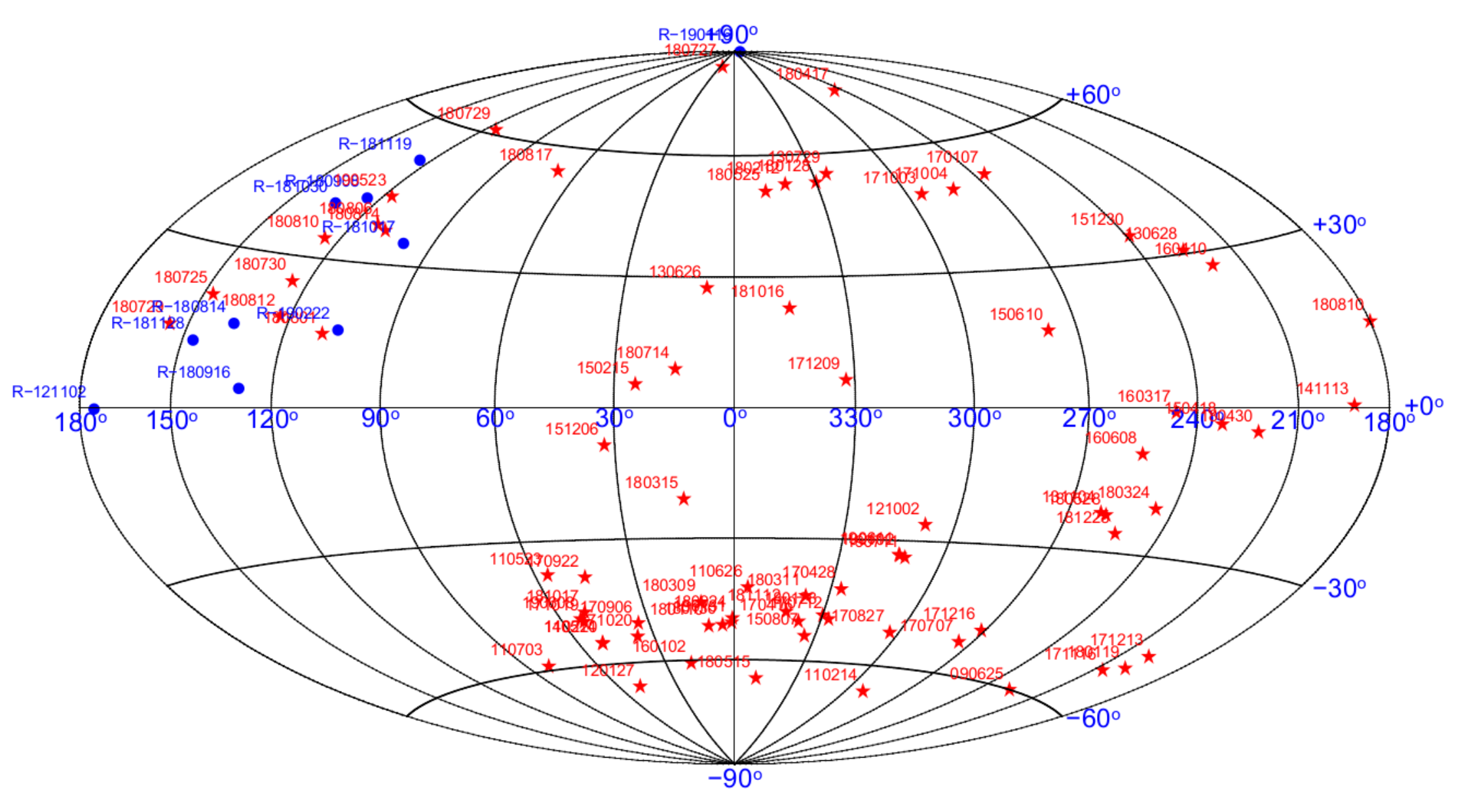}
    }
%\vspace{+0.4cm}
   \caption{The Aitoff plot of the source positions (in Galactic coordinates) included in the FRB sample selected.
    Nonrepeating source positions are marked with red stars, while those of R-FRBs are marked with blue circles.}
 \label{fig-1a}
\end{figure*}
}

%\pagebreak
\section{The {\mta FRB} Dispersion Measure Decomposition and the Case of Source 1}

{\fvv The DM of each FRB is given by the sum of the following components:}
\be \rm DM = DM_{\rm gdisk} + DM_{\rm ghalo}+ DM_{\rm host} + DM_{\rm IGM}
\label{eq-1} \en

where DM$_{\rm gdisk}$ and DM$_{\rm ghalo}$ are the Galactic disk and halo components, DM$_{\rm host}$ is the FRB source host galaxy component
from disk and halo and all specific local gas{\fvv , and DM$_{\rm IGM}$ is the intergalactic component}. When no {\fvv information} on the local host component is known, we define the excess DM as DM$_{ex} = $DM$_{\rm host} + $DM$_{\rm IGM}$. The DM$_{\rm IGM}$ has a strong distance dependence, {\fvv and} DM$_{\rm host}$ could have a correlation with age \citep[][]{2020Natur581...391}. The recently discovered \chf and ASKAP FRBs include some source{\fvv s} having small {\fvv total} DM values
%in particular, very small expected DM$_{ex}$ {\fvv with respect to the}
{\fvv only slightly in excess of the galactic contribution considering the} available Galactic DM models \cite[][]{NE2001,2019Prochaska}. Moreover, the recent precise localizations {\fvv of a few FRBs \citep[][]{2017Natur541...58,2021Univ....7...76N}  allows to} have a set of FRBs {\fvv (nine sources)} to compare the DM$_{\rm IGM}$ with the expected DM according to the intergalactic medium (IGM)-DM relation{\fvv ,} DM$\,= \,900\, z\,\rm pc\, cm^{-3}$ \citep[][]{2014McQuinn}. Source 1 has a low value of DM$_{\rm ex}$ and has been recently
localized{ \fvv with good precision by} VLBI{\fvv . It has been} associated {\fvv with} a possible star-forming region in a massive spiral galaxy at redshift $z\, =\, 0.0337\, \pm\, 0.0002$, {\fvv corresponding} to 149.0\,$\pm\,0.9$ Mpc {\fvv (M20)}.
%even due to the similar extrapolated redshift value from the measured DM.
Knowing the distance{\fvv ,} it is possible to estimate the DM$_{\rm host}$ {\fvv  for this source}. After subtracting the DM$_{\rm gdisk}$ value and assuming a value of 50\, pc\,cm$^{-3}$ for DM$_{\rm ghalo}$, we obtain DM$_{\rm ex}\,\sim\,100\,\rm pc\,cm^{-3}$, but from z\,=\,$0.0337$ we obtain DM$_{\rm IGM}\,\sim\, 32 \, \rm pc\,cm^{-3}$ and consequently DM$_{\rm host} \sim 68\,\rm pc\,cm^{-3}$ \citep[see][]{TavaniNat2020}. This source has the lowest DM$_{\rm IGM}$ among those with a certain localization.

\section{The AGILE Instrument}

AGILE is a small space mission of the Italian Space Agency (ASI) devoted to X--ray and $\gamma$-ray astrophysics \cite[][]{2009A&A...502..995T},
operating since 2007 in an equatorial orbit.
Its spinning operational mode expose{\cas s} {\fvv $\sim 80\%$ of the entire sky every 7 minutes}.
 The instrument consists of four detectors: an imaging $\gamma$-ray Silicon
Tracker \citep[sensitive in the energy range 30 MeV\,--\,30 GeV{\fvv ;}][]{2002NIMPA.490..146B},
a coded mask X-ray imager, Super-AGILE \citep[Super-A; operating in the energy range 18\,--\,60 keV][]{2007NIMPA.581..728F}, the
Mini-Calorimeter \citep[MCAL; working in the range 0.35\,--\,100 MeV;][]{2009NIMPA.598..470L},
with a 4{\fvv $\pi$} acceptance, and the anticoincidence \citep[AC;][]{2006NIMPA.556..228P} system. % completes the instrument.
The combination of Tracker, MCAL, and AC working as a $\gamma$-ray imager constitutes the AGILE/GRID.
The GRID has a field of view (FOV) of {\fvv about}
70$^{\circ}$ around {\fvv the} pointing direction, while Super-A has a 2D-coded {\fvv squared} FOV {\fvv of size $\sim\,68^{\circ}$}.
{\fvv The AGILE instrument has important characteristics for short transient follow-up
in X{\fvv -rays} and $\gamma$-rays \cite[][]{2013NuPhS.239..104P,2014ApJ...781...19B,2019ExA....48..199B,2019RLSFN...33}.}
The instrument is capable of detecting $\gamma$-ray transients and GRB-like phenomena on
timescales ranging from sub-milliseconds to tens to hundreds of seconds.
For a summary of the AGILE mission features, see \citet[][]{2019RLSFN...38}. %\\

%\section{Gamma-ray observations of FRBs}
\section{AGILE Observations of Nonrepeating FRBs}

%We investigated the possible
We considered a sample of {\fvv 89} FRB sources selected from the public catalog
FRBCAT ({\fvv consisting of} 110 sources as of 2020 May 1) {\fvv and from the \chf telescope public online database, among the verified events having a reliable position measure}.
We report the source list in Table~{\ref{tab:tab-1}}, showing the main burst radio parameters, such as the detection time (T$_0$), sky position, fluence, and measured DM{\fvr . We show in Figure~\ref{fig-1a} the plot of the source positions in Aitoff projection}. {\fvv Our main goal in this work is to focus on nonrepeating sources, and for R-FRBs we report in Table~{\ref{tab:tab-1}} the parameters of the first repetition only; a complete analysis of all the repetitions is not included in this work.}
The AGILE satellite pointing direction rotates
%{\fvv with a period of} $\sim 7$ minutes
around the axis pointed towards the Sun; so the GRID and Super-A can obtain exposures of each FRB error circle at each satellite revolution not affected by Earth occultation or SAA passages {\fvv (}MCAL is only affected by  Earth occultations{\fvv ; in the following we will indicate the unocculted region of the sky with "MCAL FOV")}.
{\fvr We applied a procedure divided in two main steps: a first check of the eventual prompt emission, supposed to include a few seconds around $T_0$ taking into account the time delay due to dispersion of the radio frequency with respect to X-rays, and then a search for delayed or preceding emission on longer time scales. }
%An initial check of the prompt
%(T$_0\,\pm$\,1 s{\fvv ; the event time correction for infinite frequency does not change the results})
{\fvr The} initial check of the prompt coverage {\fvr with the three detectors}
{\fvv of a few seconds around $T_0$ to account {\fvr for} time delay due to dispersion {\fvr value and its uncertainty (in the range $10^{-3}\,$--$\,10\,\rm pc\, cm^{-3}$)} of each {\fvv un}occulted source has been executed, and we report {\fvv the event coverage} in dedicated columns {\fvv of} Table~\ref{tab:tab-1}{\fvv, {\fvr indicating} whether} each source was within each detector's FOV at burst times (marked by {\ma the ''Y'' symbol, for }''yes'').
We {\ma determine that 1{\fvv 5} FRBs were exposed at the radio event times} within the GRID {\ma FOVs}; {\fvv instead} only one {\fvr (170107)} was inside the Super-A FOV{\fvv , while 54 events were inside the unocculted part of the sky for MCAL}. In Figure~\ref{fig-1} and~\ref{fig-3}, we show Aitoff plots in Galactic coordinates of the AGILE exposures {\ma of {\fvv seven} FRB events that occurred within the GRID FOV{\fvr , in particular the one falling also within the Super-A one}.
{\ma In our analysis}, we selected time intervals preceding and following {\ma the FRB radio times of arrival} $T_0$. {\ma We extended our searches {\fvv to} time interval{\fvv s} sufficiently large to take into account the radio dispersion delays that can amount to many seconds.}
In the following{\fvv ,} we will describe analyses of two integrations timescales: \textit{(a)}
short timescales, below 1 hr; \textit{(b)} long timescales, from 1 hr to 100 days. We will report the results {\fvv with} these timescales for the two AGILE detectors, {\ma MCAL and GRID}.
\begin{figure*}[t]
%\vspace{-1cm}
   \centerline{
\includegraphics[width=0.45\textwidth, angle = 0]{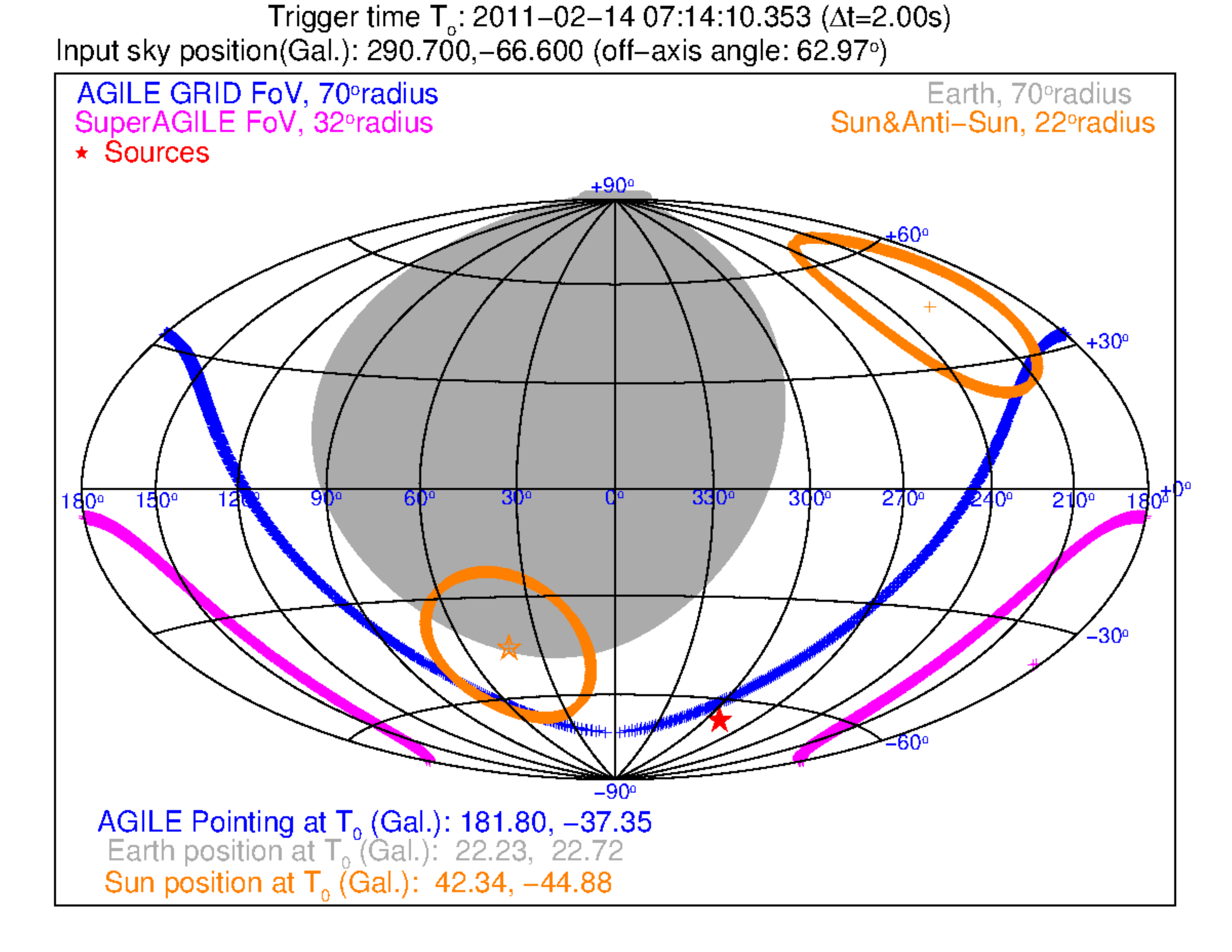}
\vspace{+0.05cm}
\includegraphics[width=0.45\textwidth, angle = 0]{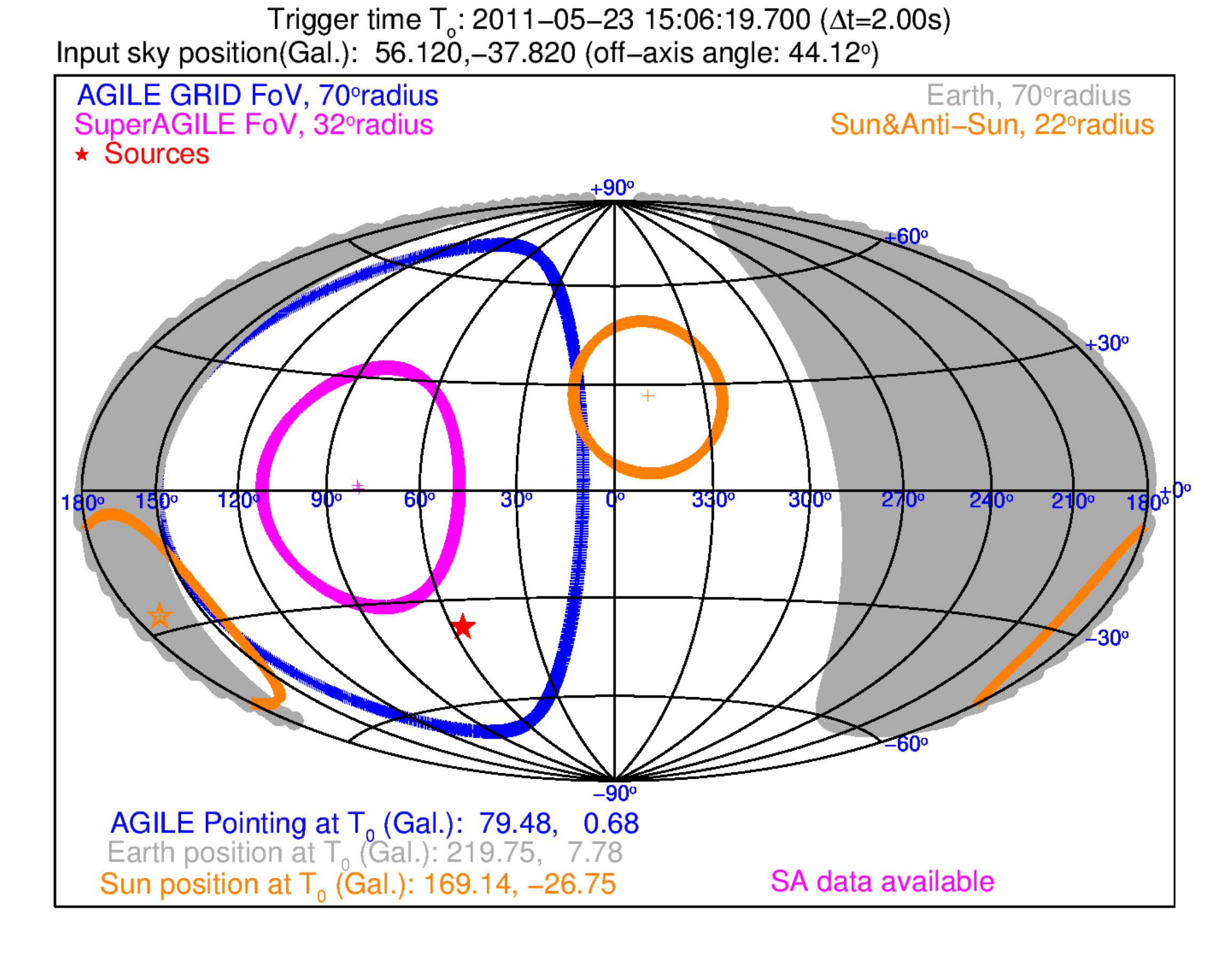}
    }
   \centerline{
\includegraphics[width=0.45\textwidth, angle = 0]{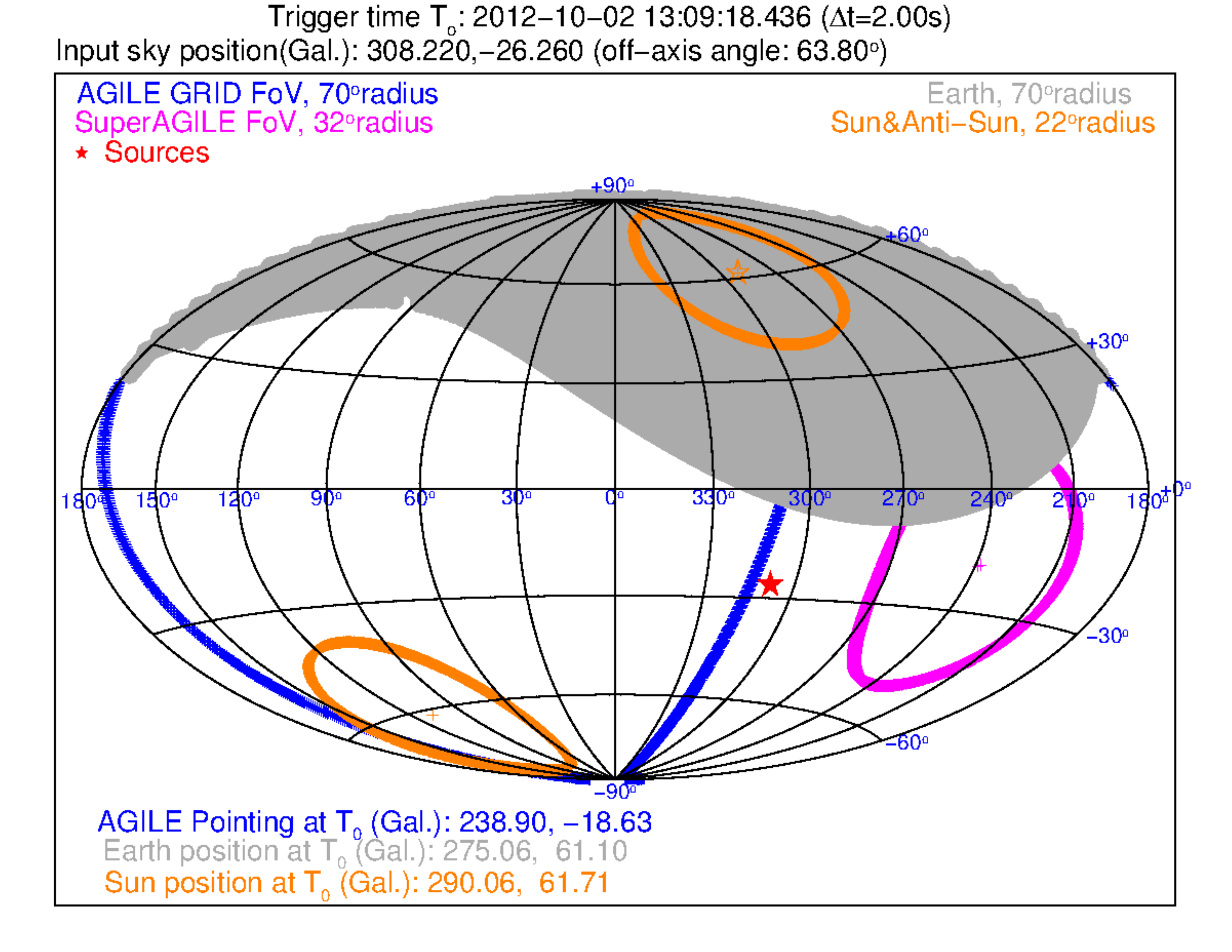}
\vspace{+0.05cm}
\includegraphics[width=0.45\textwidth, angle = 0]{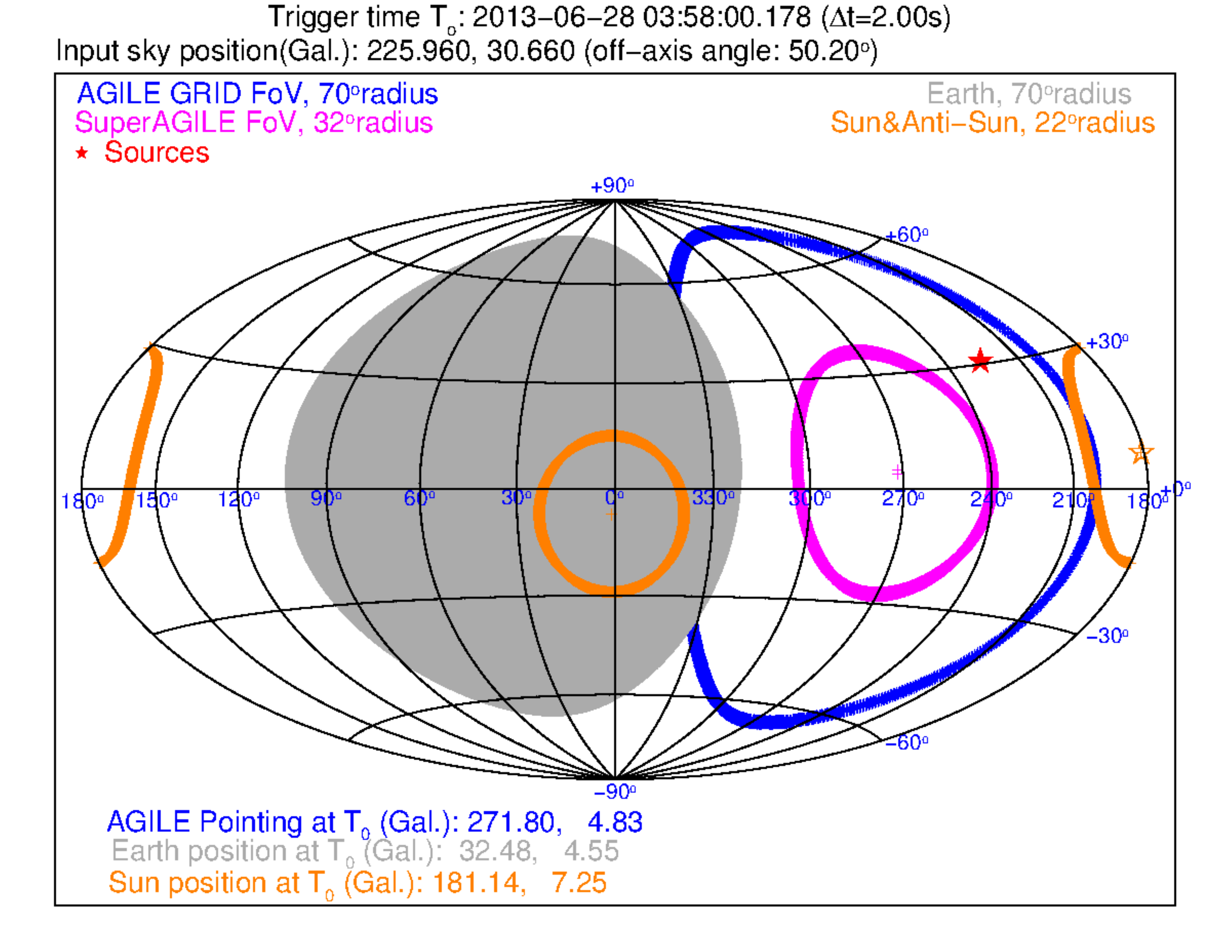}
    }
%\vspace{+0.4cm}
   \caption{Top panels: the AGILE/GRID FOV map (in Galactic coordinates) at 110214 and 110523 $T_0$ (from left to right).
    Bottom panels: the FOV map at 121002 and 130628 $T_0$.
    Source position is marked with a red star. Earth occulted region is in gray, while the Sun/anti-Sun unaccessible {\fvv (to GRID and Super-A)} regions are delimited with orange circles. The GRID FOV is shown with a blue circle, while the central part of the Super-A {\fvv FOV} is indicated with a magenta circle.}
 \label{fig-1}
\end{figure*}
%\pagebreak

%\subsection{Search for precursor and delayed gamma-ray emission}
\subsection{Search for Hard X-ray/Soft $\gamma$-ray Emission}

The AGILE/MCAL is a triggered detector, with an onboard trigger logic acting on different energy ranges and timescales (from $\sim\,300\,\mu$s to $\sim\,8$\,s). The submillisecond logic timescale allows MCAL to trigger on very short duration impulsive events. Moreover{\fvv ,} the logic is capable of efficiently detecting long and short transients such as GRBs \citep{2008A&A...490.1151M,Galli2013} as well as terrestrial $\gamma$-ray flashes on millisecond {\ma and submillisecond} timescales \citep{2011PhRvL.106a8501T,Marisaldi2015}. A detailed discussion about MCAL triggering capabilities and {\fvv UL evaluation} is given in \citet[][]{2020Casentini} and \citet[][]{2019ApJ...871...27U}.
\begin{figure*}[ht!]
%\vspace{-1cm}
   \centerline{
   \includegraphics[width=0.45\textwidth, angle = 0]{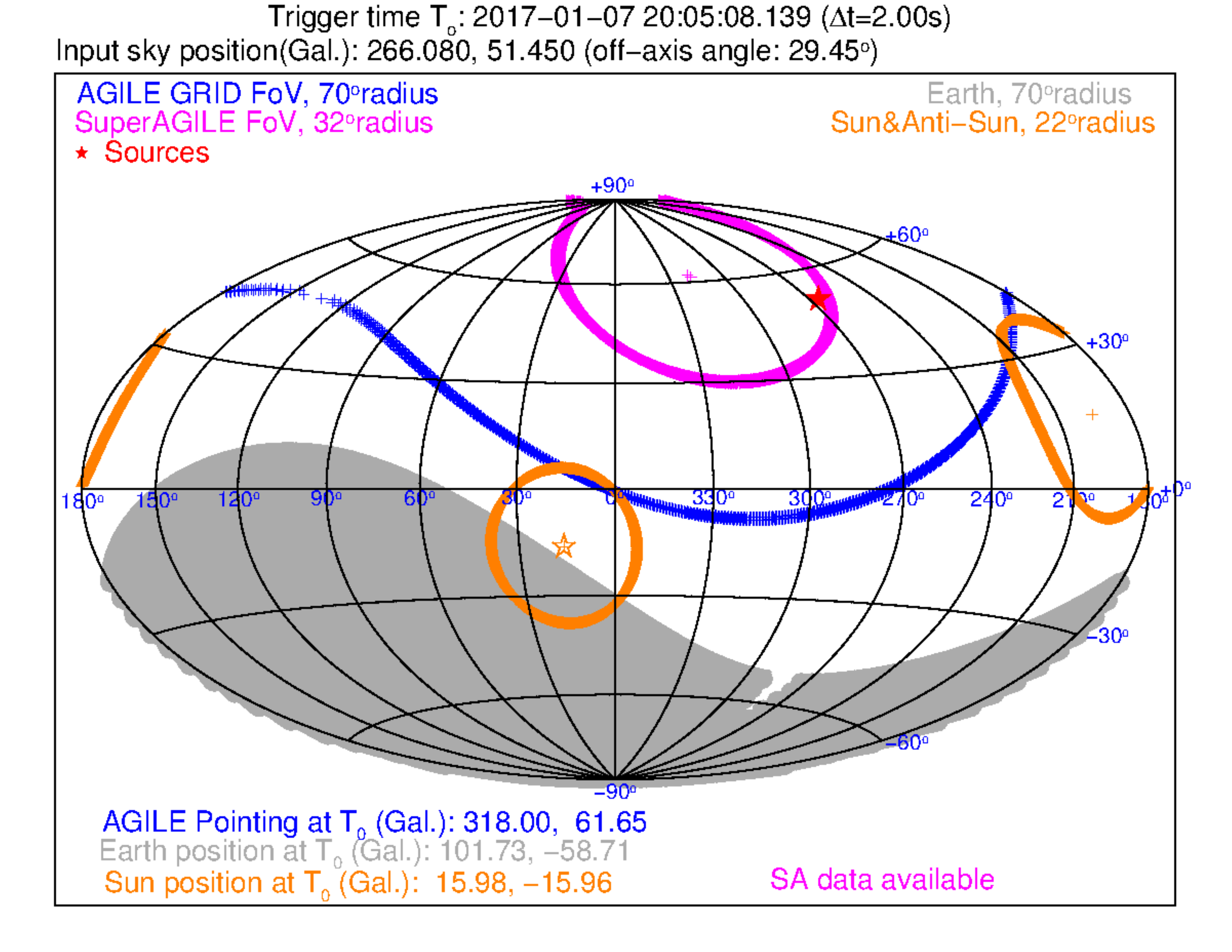}
\vspace{+0.2cm}
\includegraphics[width=0.45\textwidth, angle = 0]{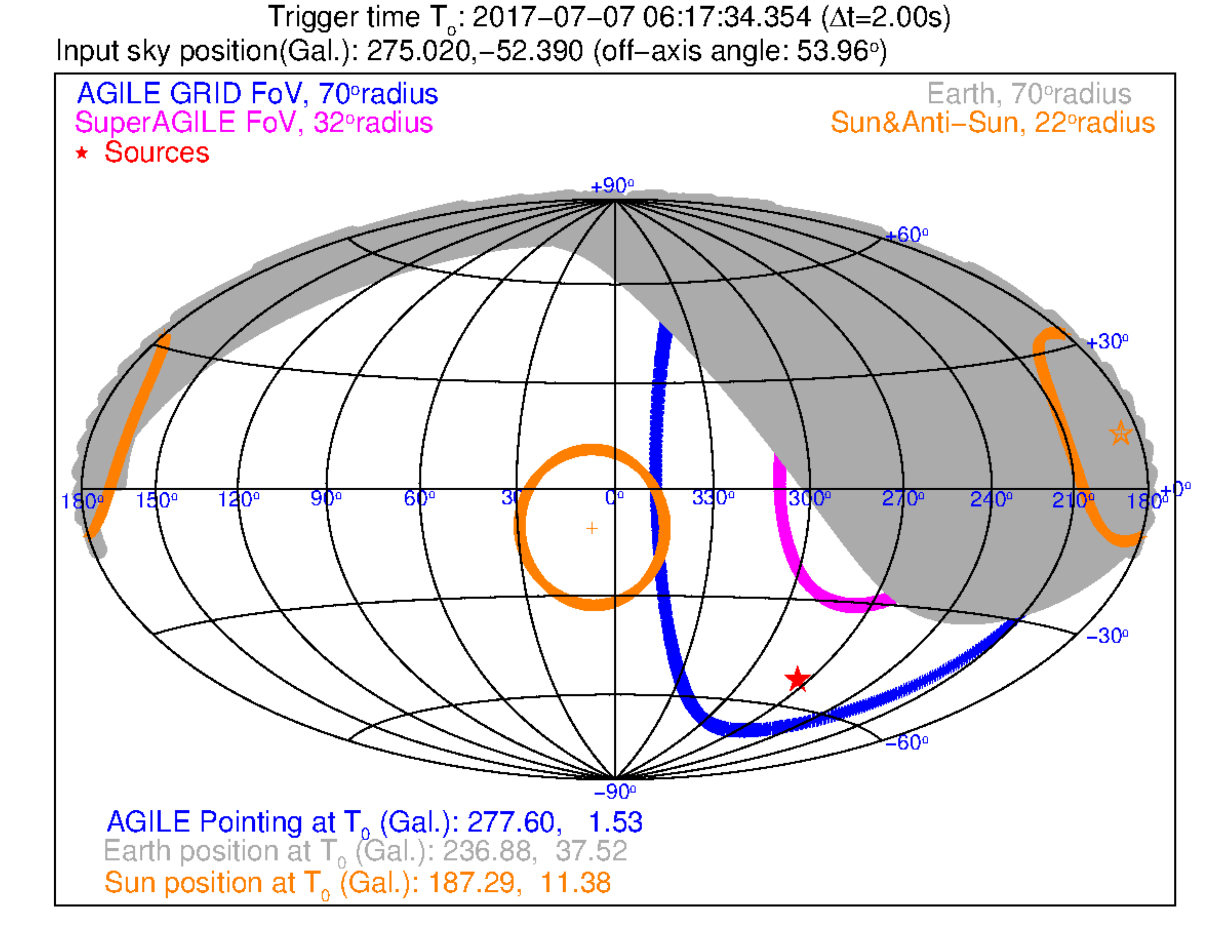}
%\vspace{+0.2cm}
    }
   \centerline{\includegraphics[width=0.45\textwidth, angle = 0]{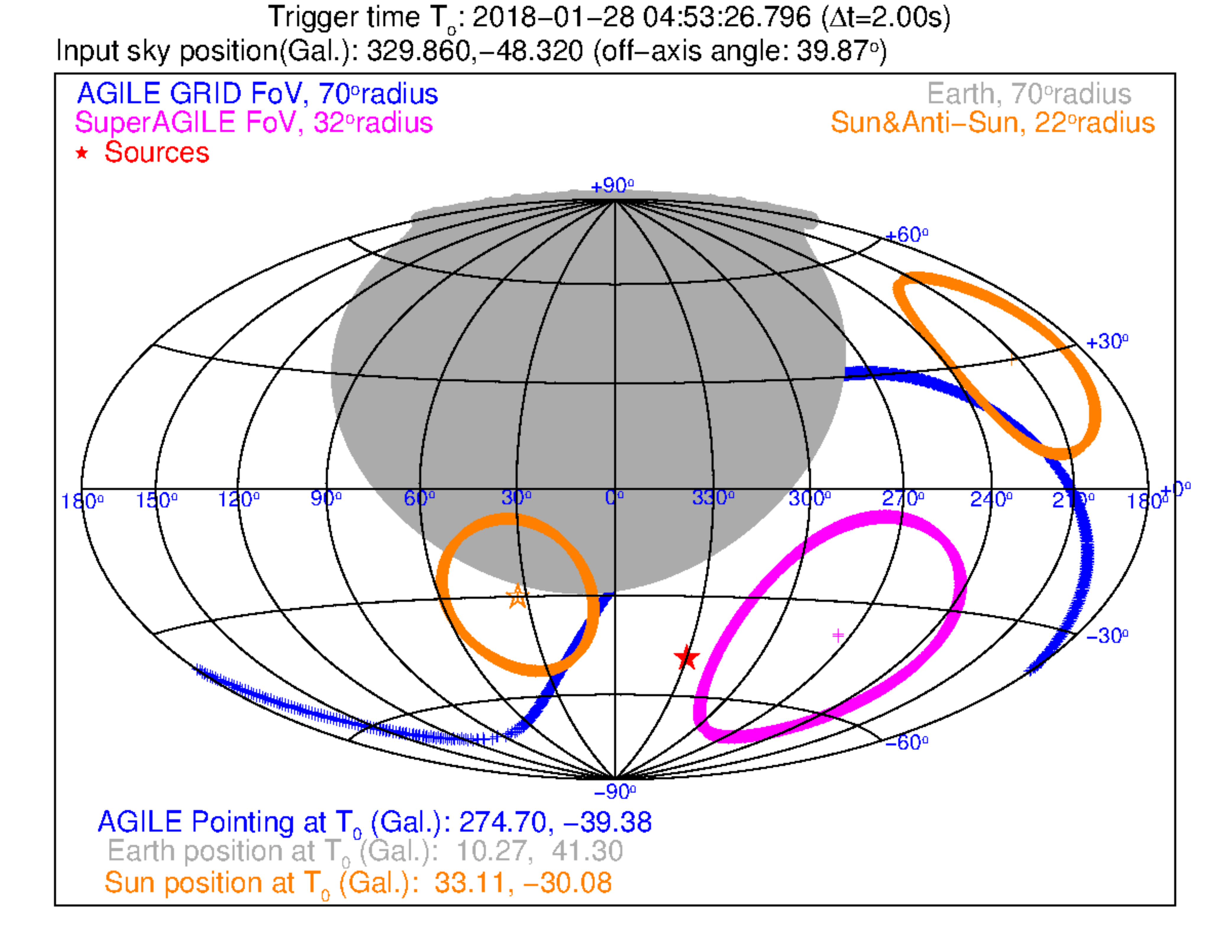}}
%\vspace{+0.4cm}
   \caption{The AGILE/GRID FOV map (in Galactic coordinates) at 170107, 170707 (from left to right) and 180128.2 $T_0$ (bottom). Source position is marked with a red star.
%   Earth occulted region is in gray, while the Sun/Anti-Sun unaccessible region are delimited with orange circles. The GRID FOV is indicated with a blue circle while the central part of the SA one is indicated with a magenta circle.
   }
 \label{fig-3}
\end{figure*}
\begin{table*}
%[ht!]
  %\begin{center}
    \caption{Typical Fluence AGILE/{\fv {MCAL}} {\fvv 3$\sigma$} Upper Limits (in {\cc {$\rm erg \, cm^{-2}$}})}
%    \hspace*{-1cm}
    \begin{tabular}{ccccccc}
     \tableline
    Sub-ms & 1 ms & 16 ms & 64 ms & 256 ms & 1024 ms & 8192 ms\\
     \tableline

     $1.13 \times 10^{-8}$ & $1.29 \times 10^{-8}$ & $3.72 \times 10^{-8}$ & $4.97 \times 10^{-8}$ & $7.95 \times 10^{-8}$ & $1.59 \times 10^{-7}$ & $4.49 \times 10^{-7}$\\

     \hline
    \end{tabular}
    \label{tab:tab-2}
 %\end{center}
\end{table*}

{\fvv As soon as} each FRB position {\fvv becomes not occulted}
%is de-occulted
and accessible by MCAL, we carried out a search for transient X{\fvv -ray} emission on the six fixed integration timescales of the trigger logic
in the 400~keV\,--\,100~MeV energy range at the data interval nearest in time{\fvv ,} within $\pm\,100 \,\rm s$.
No significant triggered or untriggered detection was obtained by MCAL.
%\subsection{Prompt emission}
We %estimated
{\ma determined} {\fvv 3$\sigma$} fluence {\fvv UL} values{\fvv ,} reported in Table \ref{tab:tab-2}, with values ranging from $1.1\, \times\, 10^{-8}\,\rm erg \, cm^{-2}${\fvv ,}
on the submillisecond timescale{\fvv ,} to $4.5\,\times\,10^{-7}\,\rm erg \, cm^{-2}$ on 8 s timescale{\fvv , using
% as spectral model a mean power-law spectrum of photon index 1.5.}
a mean power law with photon index 1.5 as the spectral model.}
 \begin{figure*}[ht!]
%% %          \vspace*{4.cm}
    \centerline{
%    \includegraphics[width=9.4cm, angle = 0]{fig7a.pdf}
%  \hspace{-0.4cm}
    \includegraphics[width=\textwidth, angle = 0]{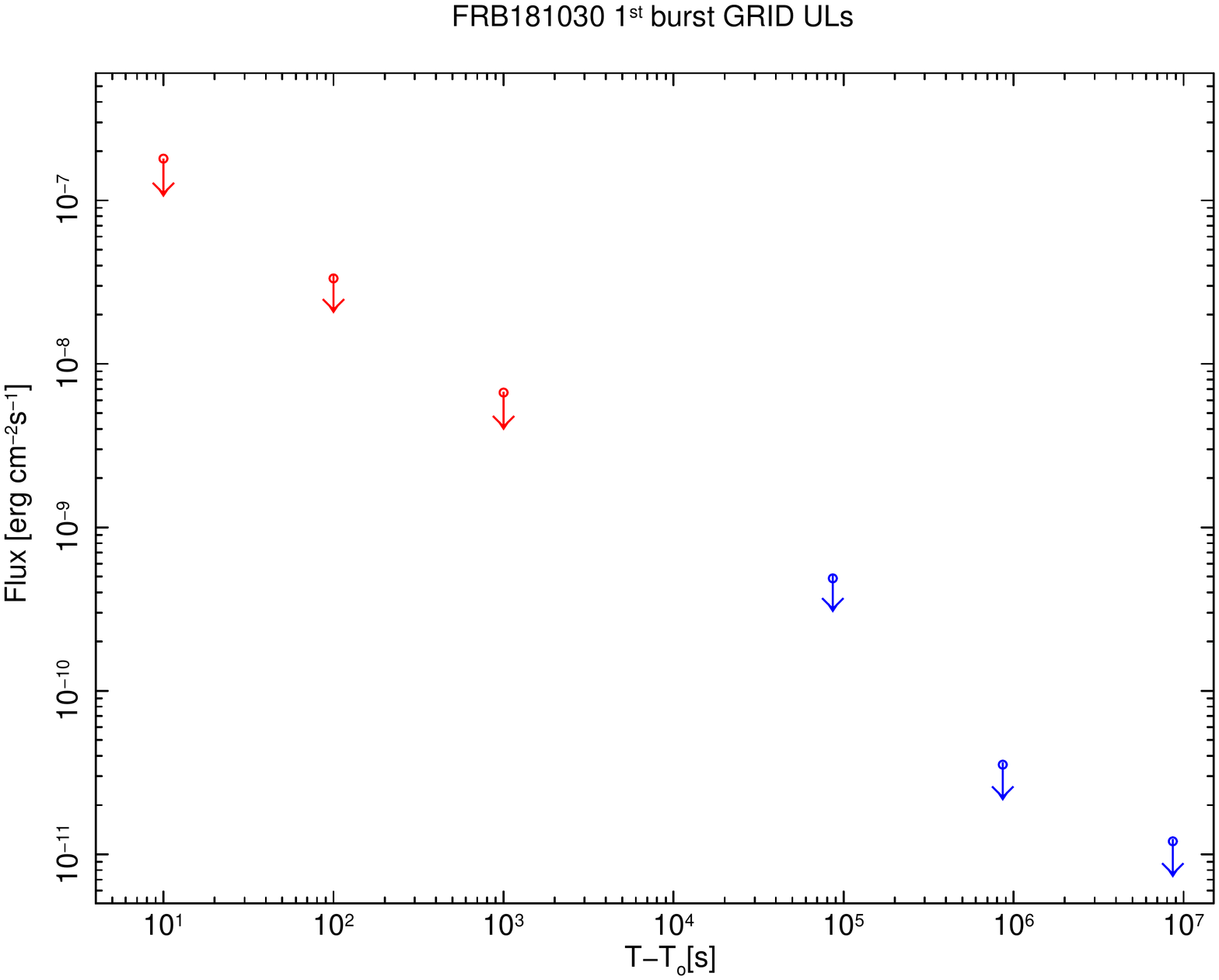}}
     \caption{The typical AGILE/GRID $\gamma$-ray 2$\sigma$ flux ULs obtained at the Source 2 position as a function of integration time. Integrations of 10, 100, and 1000 s are in the 50 MeV\,--\,10 GeV energy band {\fvv (in red)}, while those ranging from 1 to 100 days are in the 100 MeV\,--\,10 GeV band {\fvv (in blue)}.
 %On the left the ULs preceding $T_0$ till T$_0$-101 days, and on the right the ones following it till T$_0$+21 days.
 }
  \label{fig-4}
 % \vspace*{0.5cm}
 \end{figure*}
%The AGILE instrument has important characteristics for short transient follow-up
%

%\subsection{GRID data analysis}
\subsection{Search for $\gamma$-ray Emission}

We searched for $\gamma$-ray emission in the GRID detector data, applying two different types of analyses: the ``{\fvv GRB} detection mode" \cite[][]{2010ApJ...708L..84G,2017ApJ...850L...27V,2020Casentini} for short-timescale integrations of 10, 100 and 1000 s before and after the event $T_0$,
which allowed {\fvv us} to obtain flux ULs in the 50 MeV\,--\,10 GeV energy band; {\fvv and} the AGILE {\fvv multi-source} maximum likelihood (ML) analysis \cite[][]{2012Bulgarelli} for long timescales, up to 100 days in the 100 MeV\,--\,10 GeV band. {\fvv The ULs were obtained for a mean power-law spectral model with photon index 2.0.}
%Table \ref{tab-3}
%summarizes our results and upper limits for these long integrations.
%in

In the case of 110523, {\fvv with a} 100 s integration, applying the method described in \cite{1983Li&Ma},
we obtained a weak $\gamma$-ray signal spatially coincident with the FRB source at $3.2\sigma$ pretrial significance.
\begin{table*}%[h!]
%\begin{deluxetable*}{cccccc}
  \begin{center}
{\fvr     \caption{Typical Fluence AGILE/{\fv {GRID}} {\fvv 2$\sigma$} Upper Limits (in {\cc {$\rm erg \, cm^{-2}\,s^{-1}$}})}
    %\hspace*{-1cm}
    \begin{tabular}{cccccc}
%     \hline
\tableline
    10 s & 100 s & 1000 s & 1 day & 10 days & 100 days \\
%     \hline
\tableline
     $1.80 \times 10^{-7}$ & $3.33 \times 10^{-8}$ & $6.67 \times 10^{-9}$ & $4.87 \times 10^{-10}$ & $3.53 \times 10^{-11}$ & $1.20 \times 10^{-11}$ \\

     \hline
    \end{tabular}
    \label{tab:tab-3} }
 \end{center}
%\end{deluxetable*}
\end{table*}

The typical AGILE/GRID $\gamma$-ray 2$\sigma$ flux ULs for short and long integrations
are shown in Figure~\ref{fig-4}{ \fvr and Table~\ref{tab:tab-3}}, for the case of the first burst of the Source 2 repeater.

\section{Discussion}
%\label{disc}
We
%suppose the
{\ma notice that typical} FRB excess DMs {\ma are} in the range of 100\,--\,1000 \,\rm pc\,cm$^{-3}${\fvv ,}
primarily due to inter-galactic propagation of the radio signal ($\rm DM_{IGM}$; CC19),
%(e.g., Cordes and Chatterjee, 2019; hereafter CC19)
thus setting an extra-galactic scale for the FRB distances.
Typical host-galaxy contributions to the DM ($\rm DM_{host}$; as for the case of our Galaxy)
amount to values $\leq 100 \, \rm pc \, cm^{-3}$.
{\ma Therefore,} typical FRB source
extrapolated distances are  of the order of Gpc, %$d = d_{Gpc} \, \rm \, Gpc$,
except for a currently small number of FRBs with small {\ma DM} excesses
($\rm DM_{IGM} \leq 32 \, \rm pc \, cm^{-3}$) as shown in Table~\ref{tab:tab-4}. %\ref{tab:tab-3}.
{\fvv This table includes also two sources having $DM_{IGM} \,< \,100 \, \rm pc \, cm^{-3}$.}
{\ma Focusing on nearby FRB sources of Table~\ref{tab:tab-4}, we will consider the Source 1 distance as a distance {\fvv UL}
for a typical scale of
%$d = d_{150Mpc} \, \rm \, Mpc$
{\fvv $d_{150\rm Mpc}\,=\,d/(150\, \rm Mpc)$}.}
%
%---------------------------------------------
%\begin{table*}[ht!]

%\end{table*}
%%
%\begin{table*}[ht!]
%\end{table*}
% TEST!!!!!!!!!!!!!!!!!!!!!!!!!!
\begin{table*}[ht!]
\begin{center}
%\label{tab:tab-3b}
%\begin{deluxetable*}{l@{\hspace{3mm}}c@{\hspace{5mm}}c@{\hspace{5mm}}c@{\hspace{4mm}}c@{\hspace{4mm}}c@{\hspace{4mm}}c@{\hspace{5mm}}c@{\hspace{5mm}}c@{\hspace{3mm}}c@{\hspace{5mm}}c@{\hspace{3mm}}}
%{\bf Table 2: detected and extrapolated parameters of low DM$_{IGM}$ FRBs.} \vskip0.1cm
%\tablewidth{700pt}
%\tabletypesize{\scriptsize}
    \caption{Estimated parameters of {\ma FRBs with low DM$_{\rm IGM}$ (assuming DM$_{\rm ghalo}$ = 50\,$\rm pc\,cm^{-3}$)}.} \vskip0.1cm
%%\begin{tabular}{|c|c|c|c|c|c|l|}
\scriptsize{
\hspace{20mm}
%\begin{tabular}{|l@{\hspace{4mm}}|l@{\hspace{4mm}}|c@{\hspace{5mm}}|c@{\hspace{5mm}}|c@{\hspace{5mm}}|c@{\hspace{5mm}}|c@{\hspace{5mm}}|c@{\hspace{5mm}}|c@{\hspace{3mm}}|}
\begin{tabular}{l@{\hspace{3mm}}c@{\hspace{5mm}}c@{\hspace{5mm}}c@{\hspace{4mm}}c@{\hspace{4mm}}c@{\hspace{4mm}}c@{\hspace{5mm}}c@{\hspace{5mm}}c@{\hspace{3mm}}c@{\hspace{5mm}}c@{\hspace{3mm}}}
  \hline

%\begin{longrotatetable}
%\begin{deluxetable*}{lccccccccllllll}
%\tablewidth{700pt}
%\multicolumn{4}{|cccc|}{($\rm pc\,cm^{-3}$)}
%\tablehead{
% ID & DM\tablenotemark{\scriptsize{*}}  &  DM$_{gal}$ & DM$_{halo}$ & DM$_{host}$ & DM$_{IG}$\tablenotemark{\scriptsize{**}} & z$_{ext}$ & d  \\
 \colhead{ID} & \colhead{Tel.\tablenotemark{\scriptsize{1}}} & \colhead{$\rm DM$\tablenotemark{\scriptsize{2}}}  &  \colhead{$\rm DM_{\rm gal}$}  & \colhead{$\rm DM_{host}$} & \colhead{$\rm DM_{IGM}$\tablenotemark{\scriptsize{3}}} & \colhead{$z_{\rm est}$\tablenotemark{\scriptsize{4}}} & \colhead{$d_{\rm est}$\tablenotemark{\scriptsize{5}}} & \colhead{W.\tablenotemark{\scriptsize{6}}} & \colhead{{\ma AGILE}}   \\
    &  &    &   &    &  &  & &  & \colhead{Cov.\tablenotemark{\scriptsize{7}}}  \\
%    &           &    &   &    &    &   \\
%    & Start ; 2\,$\sigma$ UL &  Start ; 2\,$\sigma$ UL &  Start ; 2\,$\sigma$ UL &   \\
%    &   time  &    & \multicolumn{3}{c|}{ --------------------------- } \\
%     &  & ($\rm pc\,cm^{-3}$) & ($\rm pc\,cm^{-3}$) &  ($\rm pc\,cm^{-3}$) &  ($\rm pc\,cm^{-3}$) &    & ($\rm Mpc$) &  ($\rm ms$) &  \\
  \hline
%     &  & \multicolumn{3}{c}{($\rm pc\,cm^{-3}$)} &    &  & {\fvr($\rm Mpc$)}  & ($\rm ms$) &  \\
     &  & \multicolumn{3}{r}{($\rm pc\,cm^{-3}$)} &    &  & {\fvr($\rm Mpc$)}  & ($\rm ms$) &  \\
  \hline
%}
%\startdata
171020            & A  & 114.1\,$\pm$\,0.2    & 38.0 & $<$\,32 & $<$\,32  & $<$\,0.0340 & $<$\,150      & 3.20 & NNN  \\
171213            & A  & 158.6\,$\pm$\,0.2    & 36.0 & $<$\,41 & $<$\,32  & $<$\,0.0340 & $<$\,150      & 1.50 & YNN  \\
180430            & A  & 264.1\,$\pm$\,0.5    & 165.4 &   50   & $<$\,32  & $<$\,0.0340 & $<$\,150      & 1.20 & YNN  \\
180729.J1316+55   & C  & 109.61\,$\pm$\,0.002 & 31.0 & $<$\,32 & $<$\,32  & $<$\,0.0340 & $<$\,150      & 0.12 & YNN  \\
180810.J1159+83   & C  & 169.134\,$\pm$\,0.002& 47.0 &   50    & $<$\,32  &   0.0246    &{\ma $<$\,150} & 0.28 & YNN  \\
180814.J0422+73(R)& C  & 189.38\,$\pm$\,0.09  & 87.0 & $<$\,20 & $<$\,32  & $<$\,0.0340 & $<$\,150      &  ...  & ...   \\
\ffrbnpB(R)& C  & 103.5\,$\pm$\,0.7    & 40.0 & $<$\,32 & $<$\,32  & $<$\,0.0340 & $<$\,150      &  ...  & ...   \\
  \hline
\ffrbnp(R)& C  & 348.8\,$\pm$\,0.2    & 200.0&  68  & 32.0\tablenotemark{\scriptsize{+}}  & 0.0337\tablenotemark{\scriptsize{+}} & 149\tablenotemark{\scriptsize{+}} & ...  & ... \\
  \hline
110214            & P  & 168.9\,$\pm$\,0.5    & 31.1 &   50    &    37.8  &   0.0420    & 177           & 1.90  & YYN  \\
170707            & A  & 235.2\,$\pm$\,0.6    & 36.0 &   50    &    99.2  &   0.1102    & 465           & 3.50  & YYN  \\
%\enddata
  \hline
\end{tabular}
}
\label{tab:tab-4}
\end{center}
%\vskip-.5cm
%\noindent (*)
%\footnotetext{
{\scriptsize{\fv{(R) ~Repeater FRBs.}}\\
{\fv{ (1) ~Radio telescopes, "C" for \chfp, "A" for ASKAP and "P" for Parkes.}}  \\
\scriptsize{\fv{(2) ~Observed dispersion measure values \citep[][]{2016PASA33..e45}.}} \\
{\fv{\scriptsize (3) ~The contribution to DM from the IGM obtained by subtracting the Galactic plane contribution (depending on the sky position) and the Galactic halo (which we assumed to be equal to 50\,$\rm pc\,cm^{-3}$; \citealt[][]{2019Prochaska}) plus host galaxy contribution (the latter value has been assumed to be equal to 50\,$\rm pc\,cm^{-3}$ except for source 1 whose distance is known). However sources with low DM have uncertain $\rm DM_{host}$ value (using the previous assumptions would bring to negative values) so we assume they should have a $\rm DM_{host}$ value lower than that of Source 1.}}  \\
{\fv{\scriptsize (4,5) ~Estimated redshift according to the IGM-DM relation in \citealt[][]{2014McQuinn}{ \fvv and the corresponding distance}.}}  \\
{\fv {\scriptsize (6) ~Radio burst scattering-corrected width.}}   \\
{\fv{\scriptsize (7) ~AGILE event spatial coverage within the three detectors' FOV{\fvv ,} according to Table~\ref{tab:tab-1} {\fvv (MCAL, GRID and SA, respectively)}, where "Y" for "Yes" and "N" for "No"{\fvv , while we do not report the coverage for R-FRB, indicated with ellipses}.}}  \\
%{\fv{\scriptsize (++) ~Known measured values of inter-galactic DM, redshift and distance for the repeater Source 1 \cite[][]{2020Natur577...190}.}}   \\
{\fv {\scriptsize (+) ~The value of $\rm DM_{IGM}$ of the repeater Source 1 is known \citep[][]{2020Natur577...190,Tavani2020}.}}   \\
}
%}
%\end{deluxetable*}
\end{table*}
%%%%
% ------------------------------------------------------------------------------------tegrations of 1000s, 12 hours, 1 and 2 days.
% ...

The observed emitted radio flux densities ($S_{\rm \nu, Jy}$) of FRB events reported in Table~\ref{tab:tab-1} have
typical values in the range 1\,--\,100 Jy, on millisecond timescales. These fluxes
can be converted to energies considering the relation $E\, =\, S_{\nu}\,\times\, DM_{IGM}^2 \times K \times \Delta t_{-3}\, \times \Delta\nu_{\rm GHz}${\fvv ,}
where $F = S_{\nu, Jy}$ is the radio flux density in Jy, $\Delta t_{-3}$ is the intrinsic temporal burst duration in units of 1 ms,
$\Delta\nu_{\rm GHz}$ is the bandwidth {\fvv in units of GHz} {\ma and $K$ is a constant for the conversion from $\rm DM_{IGM}$ to distance in parsecs.}
{\ma In Figure~\ref{fig-6a} we show the values of the quantities $F\,\times DM_{ex}^2$ (which is based on measured values of DM with the Galactic plane contribution subtracted) and of $F\,\times DM_{IGM}^2$ (which is a deduced quantity; see Table~\ref{tab:tab-4}).
{\fvv The values of $DM_{IGM}^2$ are evaluated with the assumption reported in Section~2 and $\rm DM_{host}\,=\,50\,\rm pc\,cm^{-3}$ for sources for which the distance is not known; consequently, some of the sources in Table~\ref{tab:tab-4} have negative values of this quantity.}
{\ma In Figure~\ref{fig-6}{\fvv ,} we show the observed FRB radio spectral fluences and the deduced }
 isotropic FRB radio energies. The latter quantities turn out to be in the range $\sim 10^{37}\,-\,10^{43} \, {\rm erg}$.
\begin{figure*}
          %\vspace{-1cm}
%  \includegraphics[width=0.55\textwidth, angle = 0]{fig6a.pdf}\includegraphics[width=0.55\textwidth, angle = 0]{fig6b.pdf}}
  \centerline{\includegraphics[width=1.15\textwidth, angle = 0]{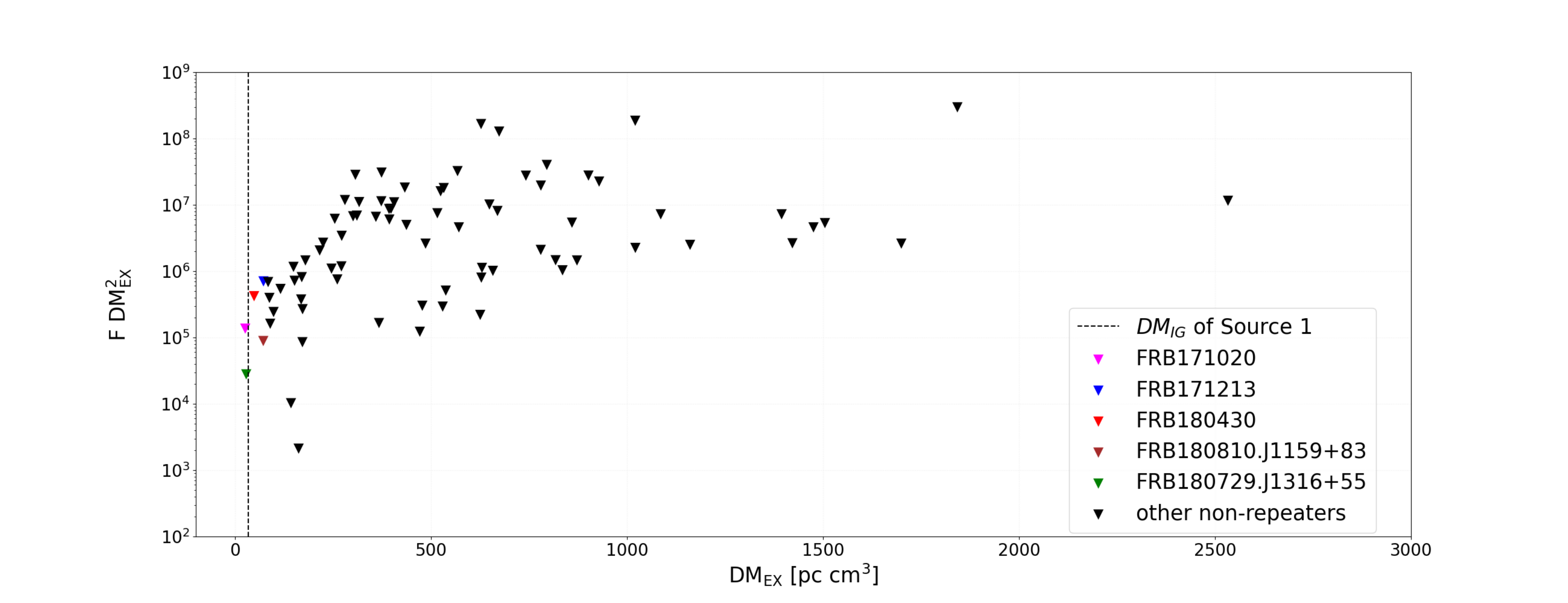}}
%  \hspace{0.2cm}
  \centerline{\includegraphics[width=1.15\textwidth, angle = 0]{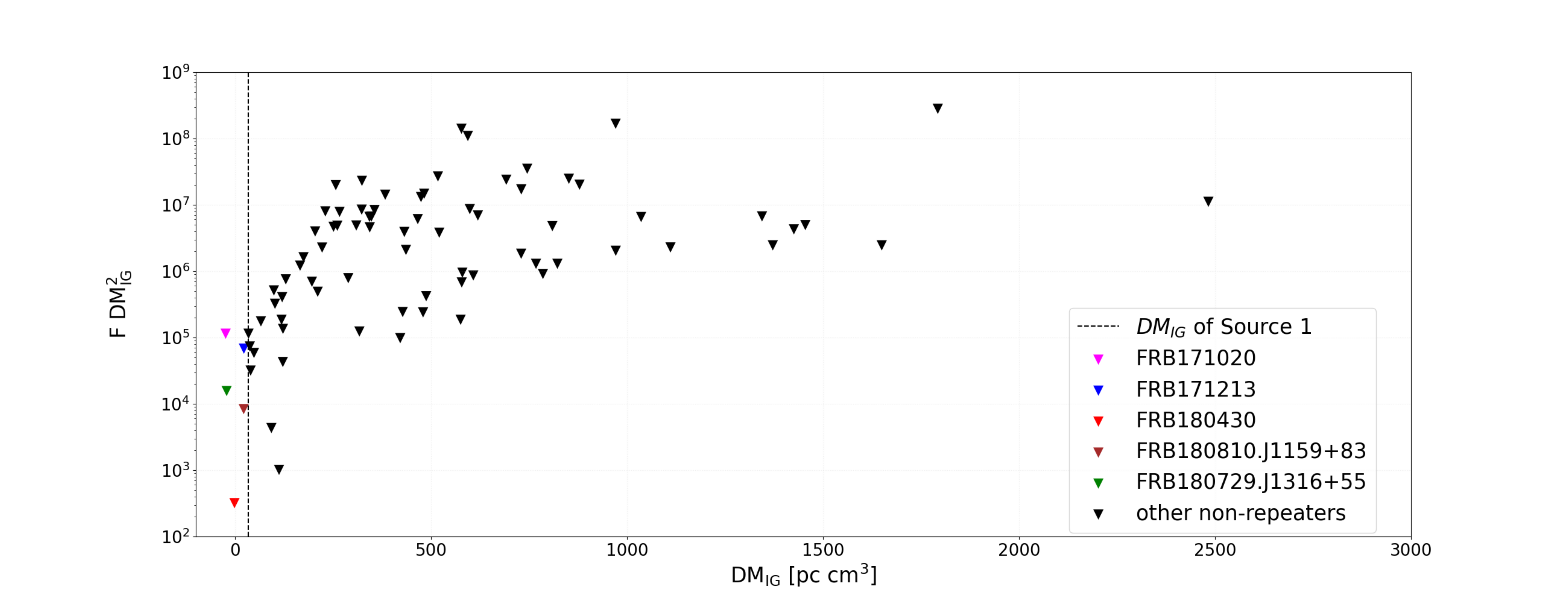}}
  \caption{ The radio flux times excess DM ($\rm DM_{ex}$; top panel) {\fvv squared} as a function of $\rm DM_{ex}$, and {\fvv radio} flux times the $\rm DM_{IGM}$  (bottom panel) {\fvv squared} as a function of $\rm DM_{IGM}$ for the nonrepeater bursts sample, where $\rm DM_{ex}$ and $\rm DM_{IGM}$ are calculated subtracting from the observed DM the Galactic disk and halo components only and the $\rm DM_{host}$ too, respectively.  }
 \label{fig-6a}
\end{figure*}

Our fluence UL $F'$ in the millisecond range at near-MeV energies has the typical value
of $F'_{\rm MeV,UL} = 10^{-8} \, \rm erg \, cm^{-2}$ {\fvv (or $10^{-7} \, \rm erg \, cm^{-2}$ in the range of seconds)} for all the FRBs exposed by AGILE/MCAL, in particular for those of Table~\ref{tab:tab-4} (except 1710{\fvv 20, which was occulted}).
This fluence {\fvv UL} translates into an UL for the isotropically MeV-radiated energy $E_{\rm MeV,UL}\, \simeq\, (2.7 \times 10^{46} \, \rm erg) \, d_{150Mpc}^2$
%  $ E_{MeV,UL} =\, 4 \, \pi F' \, d_{Gpc}^2 \simeq (10^{48} \, \rm erg) \, d_{Gpc}^2$.
%\be E_{MeV,UL} =\, 4 \, \pi F' \, d_{150Mpc}^2 \simeq ( 2.7 \times 10^{46} \, \rm erg) \, d_{150Mpc}^2
%\be E_{MeV,UL} =\, 4 \, \pi F' \, d^2 \simeq ( 2.7 \times 10^{46} \, \rm erg) \, d_{150Mpc}^2
%\label{eq-2}\en
{\fvv (or $\simeq (4.0\, \times 10^{47} \, \rm erg) \, d_{150\rm Mpc}^2$ for the second timescale). Obviously, the total emitted energy can be smaller than the isotropic equivalent when taking into account the eventual beaming effect}.
%{\ma This MeV energy UL is important {\fvv because} it is of the same order {\fvv of} the 2004 ''giant flare'' from the aGalactic magnetar \sgr \citep[][]{palmer2005,2005Natur434...1098} that displayed an emitted energy near $2\times10^{46} \, \rm erg$ during a 200 ms outburst.}
\begin{figure*}
          %\vspace{+1.0cm}
  \centerline{\includegraphics[width=1.\textwidth, angle = 0]{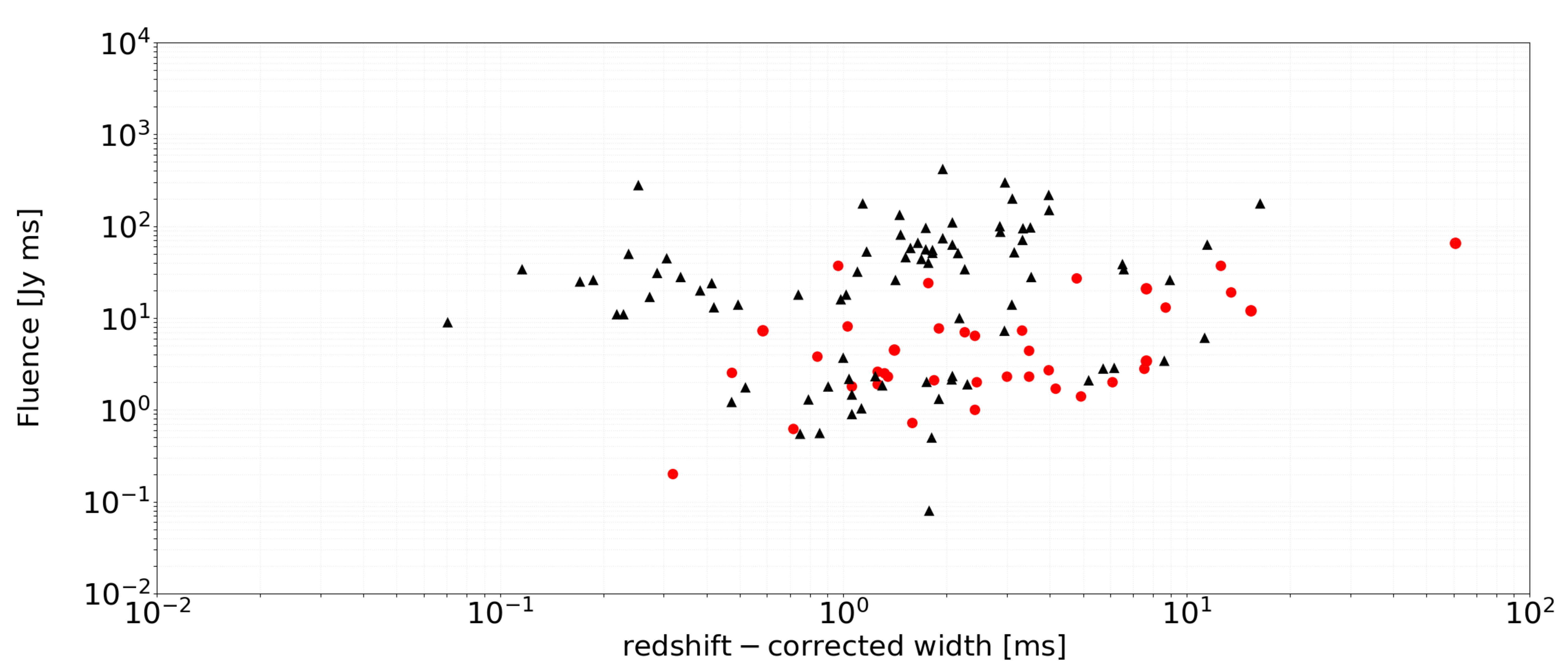}}
  \centerline{\includegraphics[width=1.\textwidth, angle = 0]{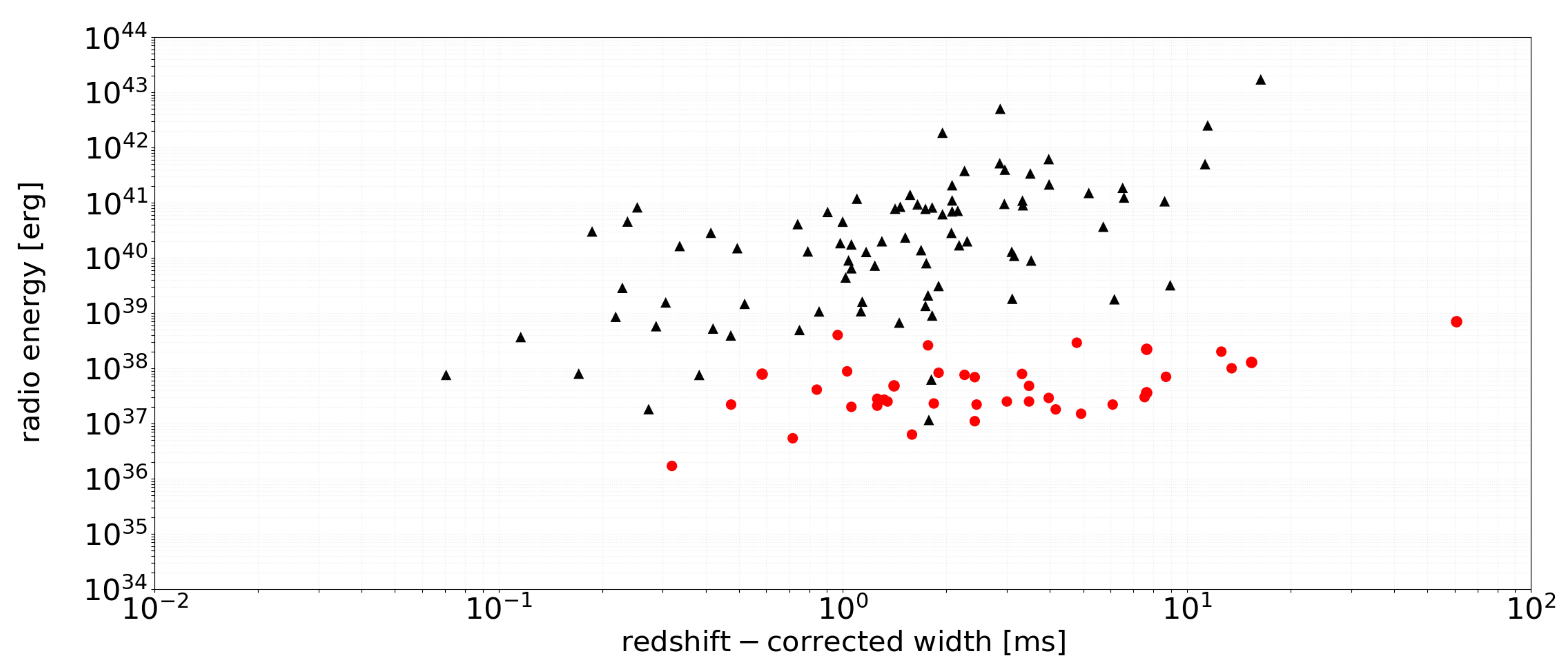}}
%  \centerline{\includegraphics[width=0.55\textwidth, angle = 0]{fig5a.pdf}
%  \hspace{0.2cm}
%  \includegraphics[width=0.55\textwidth, angle = 0]{fig5b.pdf}}%
  \caption{ The radio fluence (top panel) and isotropically emitted energy (bottom panel) vs. the redshift-corrected time widths for the same sample of nonrepeating FRBs (black triangles) and for a sample of R-FRBs (red circles). FRB radio data (typically in the 400 MHz\,---\,1.5 GHz band) and DM information are from FRBCAT \citep[][]{2016PASA33..e45}. Distances of non localized FRBs have been derived from their intergalactic DM ($\rm DM_{IGM}$) once the Galactic disk and halo contributions have been subtracted from the observed DM{\fvv ,} together with the host galaxy one (this latter value has been assumed to be such that $\rm DM_{halo} + DM_{host} = 50 + 50 = 100\,\rm pc\, cm^{-3}$, in good agreement with the direct measurement of Source 1).
  %(Bottom panels) The excess FM (DM$_{ex}$; left panel) and inter-galactic DM (DM$_{IG}$; right panel) of the same burst sample, calculated subtracting from the observed DM the galactic and halo component only and also the DM$_{host}$.
  }
% \label{fig-7}
 \label{fig-6}
\end{figure*}

{\fvv This UL is relevant because it is of the same order of magnitude of the 200 ms long first peak of the 2004 ''giant flare'' from \sgr \citep[with an emitted energy $\sim\,3\times\,10^{46}\,\rm erg$;][]{palmer2005,2005Natur434...1098}}.
{\ma The lack of MeV detections for the FRBs of Table~\ref{tab:tab-4} translates into the absence of X-ray bursts larger than the most intense giant flare from a Galactic magnetar. Our result is very significant in light of the recent discovery of a giant radio burst from the Galactic magnetar \sgrbp \citep[][]{2020Mereghetti,2020ATel13684,2020BochenekSubm,2020ATel13681,2020ApJLScholz,2020LiSubm}. FRB-like radio pulses are therefore associable to magnetars \citep[e.g.,][]{Tavani2020,2020MRAS499..2319K}, and we might expect powerful X-ray emission in coincidence with FRBs. This expectation is based on the energetics of the radio emission from FRBs as deduced from Fig. 5 that requires a source of energy presumably larger than that available in observed Galactic magnetars{\fvv , when considering those sources having high measured DM values, so are probably distant}. This source might be a fraction of the magnetic field energy that for a neutron star of radius $R\,= 10^6 \, \rm cm$ is $E_B \sim R^3 \, B_*^2 / 6 \sim 2 \times 10^{49} \, B_{*,16}^2\,\rm erg$, where $B_{*,16}$ is the internal magnetic field in units of $10^{16} \,\rm G${\fvr , where radio emission can be triggered by neutron star magnetic field instabilities or magnetospheric particle acceleration phenomena (as discussed in \citealt{thompson1996})}. It is interesting to note that in the case of \sgr (with a magnetic field deduced from the magnetar spindown $B_{*,16} \simeq 0.2$; \citealt[][]{kaspi17}) the giant outburst flare energy of $3\,\times\, 10^{46} \, \rm erg$ corresponds to a fraction of 2.5\% of the total energy $E_B$. We{\fvv } therefore{\fvv } can expect detectable X-ray and MeV flaring activity near the UL
%of Eq. \ref{eq-2}
{\fvv value $E_{\rm MeV,UL}$,} or above{\fvv ,} from FRBs if magnetars constitute the physical sources associated with the majority of FRBs of Table~\ref{tab:tab-4}. }

{A variety of FRB models discuss the possibility of X-ray and $\gamma$-ray emission from magnetar sources associated with coherent radio pulses (see CC19{\fvr , \citealt[][]{2019A&ARv27..4P} and \citealt[][]{2019PhR821..1P}} for a 2019 summary).}
The synchrotron maser modeling of FRB radio flares
% one of the models assuming a magnetar progenitor,
has been recently applied to the newly discovered \chf burst repeaters \citep[][]{2020MNRAS...494.4627M} and to the comparison of FRBs with
the \sgrbp {\fvv 1} millisecond radio flare detected on April 28$^{th}$ \citep[][]{2020ATel13681,2020ATel13684,2020ApJLScholz,2020Mereghetti,2020BochenekSubm,2020LiSubm} and {\ma with other} magnetar X-ray bursts \citep[][]{2020MargApJLSubmitt}.
%\begin{figure}
%
% \label{fig-8}
%\end{figure}
The simultaneous radio and X-ray burst detection{\fvv s} of \sgrbp allow us to obtain the radio-to-X-ray fluence ratio $E_{\rm radio}\,/\,E_X \,\sim\,10^{-5}${\fvv ,} where
$E_{\rm radio}\,\sim\,2\times\,10^{35} \,\rm erg$ and $E_{X}\,\sim\,8\times\,10^{39} \,\rm erg$ {\ma ({\fvv with the assumption of} a distance of 10 kpc, \citealt[][]{TavaniNat2020}{\fvv )}}. {\ma It is a single event but indicative of a physical phenomenon that has to produce simultaneous X-ray emission, together with unabsorbed radio pulses.}
The decelerating maser blast wave model \citep[][]{beloborodov17,2019MNRAS.485.4091M}, based on the observed parameters of the radio burst alone, {\ma leads {\fvv us} to} infer the intrinsic energy of the \sgrbp flare to be $E_{\rm flare} \sim 10^{40}\, \rm {erg}\, (\xi\,/\,10^{-3})^{-4/5} (W/5 ms)^{-1/5} d^2_{10}${\fvv ,} where $\xi$ is the intrinsic synchrotron maser efficiency{\fvv ,} supposed to be $\,\sim\, 10^{-2}\,$--$\,10^{-3}$ (for moderate magnetization), $d_{10}\, = \, d/10\,\rm kpc$, and W is the 1.4 GHz burst duration. This energy is $E_{\rm flare}\,\simeq\,E_X$ and the {\ma expected} $E_{\rm radio}\,/\,E_{\rm flare}$ {\fvv is} of the same order {\fvv as} the observed one, 10$^{-5}$.
Moreover, {\ma delayed X-ray {\mta radiation} is expected:} the ``afterglow" peak frequency is set by {\mta synchrotron emission,} and{\fvv } in the model of a decelerating ultrarelativistic blastwave{\fvv ,}
it corresponds to
$E_{\rm peak}\,\sim\,h\nu_{\rm sync}\, \simeq 160\,{\rm keV}\,\sigma^{1/2}_{-1}\,(W/5 {\rm ms})^{1/10} (t/5 {\rm ms})^{-3/2} d^2_{10}$
%\label{eq-3}
and so $E_{\rm peak}\,=\,10 \,$--$\, 100\,\rm keV$, with $t$ the time from the flare peak and $\sigma\, =\,10^{-1}\,\sigma_{-1}$ the medium magnetization parameter \citep[][]{2019MNRAS.485.4091M}. {\mta We consider} a fast cooling synchrotron spectrum with an exponential cutoff at $\nu_{\rm sync}$, so that $F_{\rm peak}\,\propto \,E_{\rm flare}$, where $F_{\rm peak}$ is the fluence at $E_{\rm peak}$. This range {\fvv of} emitted radiation is compatible with the {\ma observations of \sgrb. However, this model predicts a level of prompt X-ray emission significantly lower than the giant flare of \sgrp, and {\fvv consequently}, it does not predict detectable high-energy emission from the FRBs of Table~\ref{tab:tab-4}}.

{\ma Prompt and delayed $\gamma$-ray emission can be considered as well. A deceleration blast wave model leads to} $E_{\rm peak} \,\geq\, \rm MeV-GeV$ for  millisecond timescale emissions in the $\gamma$-ray band \citep[][]{2020MargApJLSubmitt}; these expected short bursts are different from the giant Galactic magnetar soft $\gamma$-ray flares {\ma lasting several hundreds of milliseconds}.
% Moreover, according to
% {\ma this} model, a preceding relativistic beamed radio emission with millisecond duration is expected, while an isotropic emission is detected in the giant flares, such as the 2004 flare of \sgr \citep[][]{palmer2005}.
The synchrotron maser model thus predicts that radio flares {\ma might} be followed by emissions also in the $\gamma$-ray band; the fluences (in MeV\,--\,GeV band) {\ma calculated for
SGR 1935+2154 } should be of the order of 10$^{-12}\,\rm erg\,cm^{-2}$ \citep[][]{2020MNRAS...494.4627M}.
%Due to the cut-off at $\nu_{sync}$ in the X-rays, the X-ray fluence is proportional to the E$_{flare}$ and from previous equation a correlation is inferred in by \citealtp[][]{2020MargApJLSubmitt} between observables, E$_{peak}\,\propto F_X^{1/2}t^{3/2}_X$, so higher flare energies should correspond to higher E$_{peak}$.
%
%
The FRB phenomenon {\ma might lead to high-energy emission on long} timescales, from days to months/years.
In this hypothesis AGILE/GRID observations of the FRBs of Table 1 (marked by ``YES" in the GRID column)
allowed {\ma us} to set $\gamma$-ray {\fvv UL}s at different integration timescales ranging from
tens of seconds up to hundreds of days. The longest timescale flux ULs for
100-day integrations are of the order of $\sim (1-2)\,\times\, 10^{-11} \rm \,erg\,cm^{-2}\, s^{-1}$
above 100 MeV.
For FRBs with small excess DM, this UL corresponds to the isotropic luminosity limit
%\be L_{\gamma,UL} \simeq  (1-2)\cdot 10^{45} \rm \, d_{Gpc}^2 \, erg \, s^{-1}.
\be L_{\gamma,UL} \simeq  (2-5)\,\times\, 10^{43} \rm \, d_{150\,\rm Mpc}^2 \, erg \, s^{-1}.
\label{eq-4} \en
Eq. \ref{eq-4} indicates that delayed $\gamma$-ray emission from FRBs is unlikely to be detected also for distances smaller by one order of magnitude compared to Source 1.

\section{Conclusions}

{\mta In this paper we report the results of a systematic search for high-energy emission {\fvv in AGILE data} from FRBs.}
{\fvr Outflows are expected to be leptonic; hadrons, if any, may lead to neutrino emission in case of near-magnetar or interstellar shocks, but we do not believe this is the case under the circumstances of situations leading to strong radio pulses.}
Our search for MeV and GeV emission associated with FRBs did not produce a positive detection among the {\fvv 89} FRBs that {\mta were accessible to our investigation}. {\mta Interestingly,} for a sub{\fvv sample} of these FRBs,
%{\mta associable to nearby sources}
 1{\fvv 5} events, we could determine the lack of detectable emission in the MeV range at a level that is close to the giant flare emission from the magnetar \sgrp.
Our {\fvv UL}s are particularly relevant for FRB sources
{\ma {\fvv with low intergalactic DM, i.e.,} presumably located at relatively small distances (smaller than 150 Mpc)}, whose parameters are shown in Table~\ref{tab:tab-4}.
%
%The MCAL MeV fluence ULs for millisecond timescales \citep[][]{2020Casentini}
%allow to exclude flare energies comparable with those observed in the MeV range from the 2004 \sgr \cite[][]{palmer2005}.
Currently, the distance out {\fvv to} which we can exclude an MeV flare in the millisecond timescale range similar to \sgr {\fvv for FRBs} is 100 Mpc.
Furthermore, we obtain long-timescale $\gamma$-ray luminosity limits reaching the relevant value of $\sim 10^{43} \rm \, d_{150\,Mpc}^2 \, erg \, s^{-1}$.

{\ma {\fvv Recently, {\mta other} high-energy counterpart searches for FRBs have been reported, {\mta with results qualitatively similar to those reported here} \citep[see][]{2019ApJ879...40,2019A&A...631A..62M,2020Natur577...190,2020A&A...637A..69G}, such as isotropic $\gamma$-ray luminosity limits $\geq\,10^{46}\, \rm erg \, s^{-1}$ at 1 ms timescale.}}
The search for X-ray, as well as MeV and GeV, emissions from FRBs should continue as more events are discovered and, hopefully, smaller distances are applicable.

{\mta The recent breakthrough discovery of a giant double-peaked radio burst} from {\mta the Galactic magnetar {\sgrbp} \citep[][]{2020ApJLScholz,2020BochenekSubm} was announced {\fvv when this work was in its final form}.}
% when this paper was already in its final form.}
% is {\mta very} encouraging.}
{\mta The detection of a structured X-ray burst associated with the giant radio burst from \sgrbp
\citep[][]{2020Mereghetti,TavaniNat2020,2020RidnaiaK-W,2020LiSubm} confirms that high-energy emission can be associated with coherent submillisecond radio bursts from a magnetar. These data are of crucial relevance for resolving the question of the ultimate origin of a class of, if not all, FRBs. Future investigations on the possible link between FRB sources and highly magnetized neutron stars will be presented elsewhere.}

\vspace{1.in}

{\bf Acknowledgments:}

This investigation was carried out with partial support by the ASI grant
no. I/028/12/05. We would like to acknowledge the financial support of ASI
under contract to INAF: ASI 2014-049-R.0 to ASI-SSDC. We thank the anonymous referee for the
useful comments that helped to improve the manuscript.
F.V. dedicates this work to his late father Giorgio.
% for the continous encouragement of a teenage dream.
%thanks for your continuous spiritual support to a teenage dream.
%} \\\\
%FV strongly suggests the reader the importance of assuming vaccines against SARS-CoV2,
% whichever, whenever it is possible.

%\newpage
%\bibliography{FRB_cat_procv04}
%\bibliography{FRB_cat_procov}

\begin{thebibliography}{35}
   %\begin{thebibliography}{35}
\bibliographystyle{aasjournal}
%\expandafter\ifx\csname natexlab\endcsname\relax\def\natexlab#1{#1}\fi

%% \bibitem[{{Aasi} {et~al.}(2016){Aasi}, \& et~al.}]{2017PhRvD..074001A}
%% {Aasi}, J., {et al.} (LIGO Scientific Collaboration), 2015, Class. Quantum Grav.32, 074001, 1411.4547.
%%
%% \bibitem[{{Acernese} {et~al.}(2015){Acernese}, \& et~al.}]{2017PhRvD..024001A}
%% {Acernese}, F., {et al.} (Virgo Collaboration), 2015, Class. Quantum Grav.32, 024001, 1408.3978.
%%
%% \bibitem[{{Abbott} {et~al.}(2016{\natexlab{a}}){Abbott}, {Abbott}, {Abbott},
%%   {Abernathy}, {Acernese}, {Ackley}, {Adams}, {Adams}, {Addesso}, {Adhikari},
%%   \& et~al.}]{2016PhRvD..93l2003A}
%% {Abbott}, B.~P., {Abbott}, R., {Abbott}, T.~D., {et~al.} 2016{\natexlab{a}},
%%   Physical Review D, 93, 122003
%%
%% \bibitem[{{Abbott} {et~al.}(2016{\natexlab{b}}){Abbott}, {Abbott}, {Abbott},
%%   {Abernathy}, {Acernese}, {Ackley}, {Adams}, {Adams}, {Addesso}, {Adhikari},
%%   \& et~al.}]{2016PhRvL.116m1103A}
%% {Abbott}, B.~P., {Abbott}, R., {Abbott}, T.~D., {et~al.} 2016{\natexlab{b}},
%%   Physical Review Letters, 116, 131103
%%
%% \bibitem[{{Abbott} {et~al.}(2016{\natexlab{c}}){Abbott}, {Abbott}, {Abbott},
%%   {Abernathy}, {Acernese}, {Ackley}, {Adams}, {Adams}, {Addesso}, {Adhikari},
%%   \& et~al.}]{2016PhRvL.116x1103A}
%% {Abbott}, B.~P., {Abbott}, R., {Abbott}, T.~D., {et~al.} 2016{\natexlab{c}},
%%   Physical Review Letters, 116, 241103
%%
%% \bibitem[{{Abbott} {et~al.}(2016{\natexlab{d}}){Abbott}, {Abbott}, {Abbott},
%%   {Abernathy}, {Acernese}, {Ackley}, {Adams}, {Adams}, {Addesso}, {Adhikari},
%%   \& et~al.}]{2016PhRvL.116f1102A}
%% {Abbott}, B.~P., {Abbott}, R., {Abbott}, T.~D., {et~al.} 2016{\natexlab{d}},
%%   Physical Review Letters, 116, 061102
%%
%% \bibitem[{{Abbott} {et~al.}(2016{\natexlab{e}}){Abbott}, {Abbott}, {Abbott},
%%   {Abernathy}, {Acernese}, {Ackley}, {Adams}, {Adams}, {Addesso}, {Adhikari},
%%   \& et~al.}]{2016PhRvD..93l2004A}
%% {Abbott}, B.~P., {Abbott}, R., {Abbott}, T.~D., {et~al.} 2016{\natexlab{e}},
%%   Physical Review D, 93, 122004
%%
%% \bibitem[{{Abbott} {et~al.}(2016{\natexlab{f}}){Abbott}, {Abbott}, {Abbott},
%%   {Abernathy}, {Acernese}, {Ackley}, {Adams}, {Adams}, {Addesso}, {Adhikari},
%%   \& et~al.}]{2016PhRvL.116x1102A}
%% {Abbott}, B.~P., {Abbott}, R., {Abbott}, T.~D., {et~al.} 2016{\natexlab{f}},
%%   Physical Review Letters, 116, 241102
%%
%% \bibitem[{{Abbott} {et~al.}(2017{\natexlab{a}}){Abbott}, {Abbott}, {Abbott}, {Abernathy},
%%   {Acernese}, {Ackley}, {Adams}, {Adams}, {Addesso}, {Adhikari}, \&
%%   et~al.}]{2017PhRvL170104}
%% {Abbott}, B.~P., {Abbott}, R., {Abbott}, T.~D., {et~al.} 2017{\natexlab{a}}, Physical Review
%% Letters, 118, 221101
%% % GW170814
%% %\bibitem[{{Abbott} {et~al.}(2017{\natexlab{b}}){Abbott}, {Abbott}, {Abbott}, {Abernathy},
%% %  {Acernese}, {Ackley}, {Adams}, {Adams}, {Addesso}, {Adhikari}, \&
%% %\bibitem[{{The LIGO Scientific Collaboration and Virgo Collaboration} {et~al.}(2017{\natexlab{a}}){The LIGO %Scientific Collaboration and Virgo Collaboration}, \&
%% \bibitem[{{Abbott} {et~al.}(2017{\natexlab{b}}){Abbott}, \&
%%   {et~al.}}]{2017PhRvLaccep}
%% %The LIGO Scientific Collaboration and Virgo Collaboration, et~al., 2017{\natexlab{b}}, Physical Review Letters,
%% {Abbott}, B.~P., et~al., 2017{\natexlab{b}}, Physical Review Letters, {\fvv 119, 161101} (A17b)
%%
%% %\bibitem[{{Abbott} {et~al.}(2017{\natexlab{c}}){Abbott}, {Abbott}, {Abbott}, {Abernathy},
%% %  {Acernese}, {Ackley}, {Adams}, {Adams}, {Addesso}, {Adhikari}, \&
%% %\bibitem[{{The LIGO Scientific Collaboration and Virgo Collaboration} {et~al.}(2017{\natexlab{b}}){The LIGO Scientific Collaboration and Virgo Collaboration},  \&
%% \bibitem[{{Abbott} {et~al.}(2017{\natexlab{c}}){Abbott}, \&
%%   {et~al.}}]{2017PhRvL170814}
%% ---. 2017{\natexlab{c}}, Physical Review Letters, in press
%%
%% %\bibitem[{{The LIGO Scientific Collaboration and Virgo Collaboration} {et~al.}(2017{\natexlab{c}}){The LIGO Scientific Collaboration and Virgo Collaboration}, \&
%% \bibitem[{{Abbott} {et~al.}(2017{\natexlab{d}}){Abbott}, \&
%%   {et~al.}}]{2017PhRvLMMAsub}
%% ---. 2017{\natexlab{d}}, ApjL, in press, doi:10.3847/2041-8213/aa91c9 (MMA17)
%%
%% %\bibitem[{{The LIGO Scientific Collaboration and Virgo Collaboration} {et~al.}(2017{\natexlab{d}}){The LIGO Scientific Collaboration and Virgo Collaboration}}]{2017ApJLVC-GBM-INT}
%% \bibitem[{{Abbott} {et~al.}(2017{\natexlab{e}}){Abbott}, \&
%%   {et~al.}}]{2017ApJLVC-GBM-INT}
%% ---. 2017{\natexlab{e}}, ApJL, in preparation{\fv , doi:10.3847/2041-8213/aa920c} (LVC-GBM-INTEGRAL)


%\bibitem[{{Abdo} {et~al.}(2009)}]{2009Natur.462..331A}
%{Abdo}, A.~A., {Ackermann}, M., {Ajello}, M., {et~al.} 2009, \nat, 462, 331

%\bibitem[{{Ackermann} {et~al.}(2016){Ackermann}, {Ajello}, {Albert},
%  {Anderson}, {Arimoto}, {Atwood}, {Axelsson}, {Baldini}, {Ballet},
%  {Barbiellini}, {Baring}, {Bastieri}, {Becerra Gonzalez}, {Bellazzini},
%  {Bissaldi}, {Blandford}, {Bloom}, {Bonino}, {Bottacini}, {Brandt}, {Bregeon},
%  {Britto}, {Bruel}, {Buehler}, {Burnett}, {Buson}, {Caliandro}, {Cameron},
%  {Caputo}, {Caragiulo}, {Caraveo}, {Casandjian}, {Cavazzuti}, {Charles},
%  {Chekhtman}, {Chiang}, {Chiaro}, {Ciprini}, {Cohen-Tanugi}, {Cominsky},
%  {Condon}, {Costanza}, {Cuoco}, {Cutini}, {D'Ammando}, {de Palma}, {Desiante},
%  {Digel}, {Di Lalla}, {Di Mauro}, {Di Venere}, {Dom{\'{\i}}nguez}, {Drell},
%  {Dubois}, {Dumora}, {Favuzzi}, {Fegan}, {Ferrara}, {Franckowiak}, {Fukazawa},
%  {Funk}, {Fusco}, {Gargano}, {Gasparrini}, {Gehrels}, {Giglietto}, {Giomi},
%  {Giommi}, {Giordano}, {Giroletti}, {Glanzman}, {Godfrey}, {Gomez-Vargas},
%  {Granot}, {Green}, {Grenier}, {Grondin}, {Grove}, {Guillemot}, {Guiriec},
%  {Hadasch}, {Harding}, {Hays}, {Hewitt}, {Hill}, {Horan}, {Jogler},
%  {J{\'o}hannesson}, {Kamae}, {Kensei}, {Kocevski}, {Kuss}, {La Mura},
%  {Larsson}, {Latronico}, {Lemoine-Goumard}, {Li}, {Li}, {Longo}, {Loparco},
%  {Lovellette}, {Lubrano}, {Madejski}, {Magill}, {Maldera}, {Manfreda},
%  {Marelli}, {Mayer}, {Mazziotta}, {McEnery}, {Meyer}, {Michelson}, {Mirabal},
%  {Mizuno}, {Moiseev}, {Monzani}, {Moretti}, {Morselli}, {Moskalenko},
%  {Murgia}, {Negro}, {Nuss}, {Ohsugi}, {Omodei}, {Orienti}, {Orlando}, {Ormes},
%  {Paneque}, {Perkins}, {Pesce-Rollins}, {Piron}, {Pivato}, {Porter},
%  {Racusin}, {Rain{\`o}}, {Rando}, {Razzaque}, {Reimer}, {Reimer}, {Reposeur},
%  {Ritz}, {Rochester}, {Romani}, {Saz Parkinson}, {Sgr{\`o}}, {Simone},
%  {Siskind}, {Smith}, {Spada}, {Spandre}, {Spinelli}, {Suson}, {Tajima},
%  {Thayer}, {Thayer}, {Thompson}, {Tibaldo}, {Torres}, {Troja}, {Uchiyama},
%  {Venters}, {Vianello}, {Wood}, {Wood}, {Zaharijas}, {Zhu}, \&
%  {Zimmer}}]{2016ApJ...823L...2A}
%{Ackermann}, M., {Ajello}, M., {Albert}, A., {et~al.} 2016, ApJL, 823, L2

% EM


\bibitem[{Barbiellini} {et~al.}(2002)]{2002NIMPA.490..146B}
{Barbiellini}, G., {Fedel}, G., {Liello}, F., {et~al.} 2002, NIM A, 490, 146

\bibitem[{Beloberodov}(2017)]{beloborodov17}
{Beloborodov}, A.~M. 2017, ApJL, 843, L26

\bibitem[{Bochenek} {et~al.}(2020a)]{2020ATel13684}
{Bochenek}, C., {Kulkarni}, S., {Ravi}, V., {et~al.} 2020a, ATel, 13684, 1

% STARE2
\bibitem[{Bochenek} {et~al.}(2020b)]{2020BochenekSubm}
{Bochenek}, C., {Kulkarni}, S., {Ravi}, V., {et~al.} 2020b, Natur, 587, 59

\bibitem[{{Bulgarelli} {et~al.}(2019)}]{2019ExA....48..199B}
{Bulgarelli}, A. 2019, ExA, 48, 199

% ML
\bibitem[{{Bulgarelli} {et~al.}(2012)}]{2012Bulgarelli}
{Bulgarelli}, A., {Chen}, A.W., {Tavani}, M., {et al.} 2012, \aap, 540, A79

% ASys
\bibitem[{{Bulgarelli} {et~al.}(2014)}]{2014ApJ...781...19B}
{Bulgarelli}, A., {Trifoglio}, M., {Gianotti}, F., {et~al.} 2014, ApJ, 781, 19

% AGILE GCN!
%\bibitem[{{Bulgarelli} {et~al.}(2017){Bulgarelli}, {Tavani}, {Verrecchia}, {Cardillo},
%  {Piano}, \& {et~al.}}]{2017GCN..21564...1B}
%{Bulgarelli}, A., {Tavani}, M., {Verrecchia}, F., {Cardillo}, M., {Piano}, G., {et~al.} 2017, GRB
%  Coordinates Network, 21564

%\bibitem[{{Connaughton} {et~al.}(2016){Connaughton}, {Burns}, {Goldstein},
%  {Blackburn}, {Briggs}, {Zhang}, {Camp}, {Christensen}, {Hui}, {Jenke},
%  {Littenberg}, {McEnery}, {Racusin}, {Shawhan}, {Singer}, {Veitch},
%  {Wilson-Hodge}, {Bhat}, {Bissaldi}, {Cleveland}, {Fitzpatrick}, {Giles},
%  {Gibby}, {von Kienlin}, {Kippen}, {McBreen}, {Mailyan}, {Meegan}, {Paciesas},
%  {Preece}, {Roberts}, {Sparke}, {Stanbro}, {Toelge}, \&
%  {Veres}}]{2016ApJ...826L...6C}
%{Connaughton}, V., {Burns}, E., {Goldstein}, A., {et~al.} 2016, ApJL, 826, L6
%\bibitem[{{Goldstein} {et~al.}(2017){Goldstein}, {Burns}, {Goldstein},
%  {Blackburn}, {Briggs}, {Zhang}, {Camp}, {Christensen}, {Hui}, {Jenke},
%  {Littenberg}, {McEnery}, {Racusin}, {Shawhan}, {Singer}, {Veitch},
%  {Wilson-Hodge}, {Bhat}, {Bissaldi}, {Cleveland}, {Fitzpatrick}, {Giles},
%  {Gibby}, {von Kienlin}, {Kippen}, {McBreen}, {Mailyan}, {Meegan}, {Paciesas},
%  {Preece}, {Roberts}, {Sparke}, {Stanbro}, {Toelge}, \&
%  {Veres}}]{2017ApJGoldstein}
%{Goldstein}, A., {et~al.} 2017, ApJL, submitted
% Fermi-GBM GRB 170817A
%\bibitem[{{Connaughton} {et~al.}(2017)}]{2017GCN..21506...1C}
%{Connaughton}, {Blackburn}, {Briggs}, {et~al.},
%{Connaughton}, V., {Blackburn}, L., {Briggs}, M.S., {et~al.},
%  2017, GRB Coordinates Network, 21506
%%%
\bibitem[{Casentini} {et~al.}(2020)]{2020Casentini}
{Casentini}, C., {Verrecchia}, F., {Tavani}, M., {et~al.} 2020, ApJL, 890, L32
% {\fv (si potrebbe non mettere in biblio)}

%\bibitem[{Champion} {et~al.}(2016)]{2016Champion}
%{Champion}, D.~J., {Petroff}, E., {Kramer}, M., {et~al.}, 2016, MNRAS, 460, L30

\bibitem[{Chatterjee} {et~al.}(2017)]{2017Natur541...58}
{Chatterjee}, S., {Law}, C.~J., {Wharton}, R.~S., {et al.} 2017, Natur, 541, 58

%\bibitem[{Chatterjee} {et~al.}(2020)]{2020ApJ896...L41}
%{Chatterjee}, S., {Law}, C.~J., {Wharton}, R.~S., {et al.}, 2020, ApJ, 896, L41

\bibitem[{\chf Collaboration} {et~al.}(2019a)]{2019ApJLCHIME}
%[{CHIME/FRB Collaboration} {et al.} 2019a]
{CHIME/FRB Collaboration}, {Amiri}, M., {Bandura}, K., {Bhardwaj}, M., {et~al.} 2019a, Natur, 566, 230

\bibitem[{\chf Collaboration} {et~al.}(2019b)]{2019ApJLCHIMEb}
{CHIME/FRB Collaboration}, {Amiri}, M., {Bandura}, K., {Bhardwaj}, M., {et~al.} 2019b, Natur, 566, 235

% 8 new!
\bibitem[{CHIME/FRB Collaboration} {et~al.}(2019c)]{2019ApJLCHIMEc}
%[{CHIME/FRB Collaboration} {et al.} 2019c]
{CHIME/FRB Collaboration}, {Andersen}, B.~C., {Bandura}, K., {Bhardwaj}, M., {et~al.} 2019c, ApJL, 885, 24 (C19)

% periodicity!!
\bibitem[{CHIME/FRB Collaboration} {et~al.}(2020)]{2020ApJLCHIME}
{CHIME/FRB Collaboration}, {Amiri}, M., {Andersen}, B.~C., {Bandura}, K., {Bhardwaj}, M., {et~al.} 2020, Natur, 582, 351 (C20)

%\bibitem[{{Connaughton} {et~al.}(2016)}]{2016ApJ...826L...6C}
%{Connaughton}, V., {Burns}, E., {Goldstein}, A., {et~al.} 2016, ApJL, 826, L6
%
% models
\bibitem[{Cordes} \& {Chatterjee}(2019)]{2019ARA&A119..161101}
{Cordes}, J.~M., and {Chatterjee}, S. 2019, ARA\&A, 119, 161101 (CC19), doi:10.1146/annurev-astro-091918-
104501

\bibitem[{Cordes} \& {Lazio}(2002)] {NE2001}
{Cordes}, J.~M., \& {Lazio}, T. J. W. 2002, arXiv preprint astro-ph/0207156

\bibitem[{Cunningham} {et~al.}(2019)]{2019ApJ879...40}
{Cunningham}, V., {Cenko}, S.~B., {Burns}, E., {et~al.} 2019, ApJ, 879, 40
%

%
%\bibitem[{{Evans} {et~al.}(2017{\natexlab{b}}){Evans}, \&
%  {et~al.}}]{2017Evans}
%{Evans}, P., {Cenko}, B., {Kennea}, J.A., {et~al.}, 2017{\natexlab{b}}, Science, submitted

\bibitem[{{Feroci} {et~al.}(2007)}]{2007NIMPA.581..728F}
{Feroci}, M., {Costa}, E., {Soffitta}, P., {et~al.} 2007, NIM A, 581, 728

\bibitem[{{Fonseca} {et~al.}(2020)}]{2020ApJL.891..L6F}
{Fonseca}, E., {Andersen}, B.C., {Bhardwaj}, M., {et~al.} 2020, \apjl, 891, L6

%\bibitem[{{Fuschino} {et~al.}(2008)}]{2008NIMPA.588...17F}
%{Fuschino}, F., {Labanti}, C., {Galli}, M., {et~al.} 2008, NIM A, 588, 17

\bibitem[{{Galli} {et~al.}(2013)}]{Galli2013}
{Galli}, M., {Marisaldi}, M., {Fuschino}, F., {et~al.} 2013, A\&A, 553, A33

\bibitem[{{Giuliani} {et~al.}(2010)}]{2010ApJ...708L..84G}
{Giuliani}, A., {Fuschino}, F., {Vianello}, G., {et~al.} 2010, \apjl, 708, L84

%\bibitem[{{Giuliani} {et~al.}(2013)}]{2013GCN..15479...1G}
%{Giuliani}, A., {Longo}, F., {Verrecchia}, F., {et~al.} 2013, GRB Coordinates Network, 15479

%\bibitem[{{Giuliani} {et~al.}(2008)}]{2008A&A...491L..25G}
%{Giuliani}, A., {Mereghetti}, S., {Fornari}, F., {et~al.} 2008, \aap, 491, L25
%
%\bibitem[{{Giuliani} {et~al.}(2014){Giuliani}, {Mereghetti}, {Marisaldi},
%  {Longo}, {Del Monte}, {Pittori}, {Verrecchia}, {Tavani}, {Cattaneo},
%  {Pacciani}, {Vercellone}, \& {Rappoldi}}]{2014arXiv1407.0238G}
%{Giuliani}, A., {Mereghetti}, S., {Marisaldi}, M., {et~al.} 2014, ArXiv e-prints

%\bibitem[{{Goldstein} {et~al.}(2017)}]{2017ApJGoldsub}
%{Goldstein}, A., {\fvv {Veres}, P., {Burns}, E.,} {et~al.}, 2017, ApJL, 848, in press, doi:10.3847/2041-8213/aa8f41

\bibitem[{{Guidorzi} {et~al.}(2020)}]{2020A&A...637A..69G}
{Guidorzi}, C., {Marongiu}, M., {Martone}, R., {et~al.} 2020, \aap, 637A, 69

\bibitem[{Hurley} {et al.}(2005)]{2005Natur434...1098}
{Hurley}, K., {Boggs}, S.~E., {Smith}, D.~M., {et al.} 2005, Natur, 434, 1098

\bibitem[{Kaspi}(2017)]{kaspi17}
{Kaspi},V.~M., {Beloborodov}, A.~M. 2017, \araa, 55, 261

\bibitem[{Katz}(2019)]{2019MRAS487..491K}
{Katz}, J.~I. 2019, \mnras, 487, 491

\bibitem[{Katz}(2020)]{2020MRAS499..2319K}
{Katz}, J.~I. 2020, \mnras, 499, 2319

\bibitem[{{Labanti} {et~al.}(2009)}]{2009NIMPA.598..470L}
{Labanti}, C., {Marisaldi}, M., {Fuschino}, F., {et~al.} 2009, NIMPA, 598, 470

% Insight-HXMT
\bibitem[{Li} {et~al.}(2020)]{2020LiSubm}
{Li}, C.~K., {Lin}, L., {Xion}, S.~L., {et~al.} 2020, Nature Astronomy, 5, 378
% submitted, arXiv e-prints, arXiv:2005.11071
%%%% Li&Ma
\bibitem[{{Li} \& {Ma}(1983)}]{1983Li&Ma}
{Li}, T. \& {Ma}, Y. 1983, ApJ, 272, 317


%\bibitem[{Lorimer} {et~al.}(2006)]{2006MNRAS.372.777}
%{Lorimer}, D.~R., {Faulkner}, A.~J., {Lyne}, A.~G., {et~al.}, 2006, MNRAS, 372, 777

\bibitem[{Lorimer} {et~al.}(2007)]{2007Science.218.777}
{Lorimer}, D.~R., {Bailes}, M., {McLaughlin}, M.~A., {Narkevic}, D.~J., \& {Crawford}, F. 2007, Science, 318, 777

% Li&Ma method!
%\bibitem[{{Li} \& {Ma}(1983)}]{1983Li&Ma}
%{Li}, T. \& {Ma}, Y. 1983, ApJ, 272, 317

%\bibitem[{{Longo} {et~al.}(2012)}]{2012A&A...547A..95L}
%{Longo}, F., {Moretti}, E., {Nava}, L., {et~al.} 2012, A. \&\ A., 547, A95
%%%
\bibitem[{Macquart} {et~al.}(2020)]{2020Natur581...391}
{Macquart}, J.~P., {Prochaska}, J.~X., {McQuinn}, M., {et~al.} 2020, Natur, 581, 391

%\bibitem[{Marcote} {et~al.}(2017)]{2017ApJL834...L8}
%{Marcote}, B., {Paragi}, Z., {Hessels}, J.~W.~T., {et~al.}, 2017, \apjl, 834, L8

% New on 180916
\bibitem[{Marcote} {et~al.}(2020)]{2020Natur577...190}
{Marcote}, B, {Nimmo}, K., {Hessels}, J.~W.~T., {et~al.} 2020, Natur, 577, 190 %(M20)

\bibitem[{{Margalit} {et~al.}(2020a)}]{2020MargApJLSubmitt}
{Margalit}, B., {Beniamini}, P., {Sridhar}, N., {et~al.} 2020a, \apjl, 899, L27
%submitted, arXiv e-prints, arXiv:2005.05283

\bibitem[{{Margalit} {et~al.}(2020b)}]{2020MNRAS...494.4627M}
{Margalit}, B., {Metzger}, B.~D., {Sironi}, L., {et~al.} 2020b, \mnras, 494, 4627

%\bibitem[{{Marisaldi} {et~al.}(2014)}]{2014JGRA..119.1337M}
%{Marisaldi}, M., {Fuschino}, F., {Tavani}, M., {et~al.} 2014, Journal of
%  Geophysical Research (Space Physics), 119, 1337

\bibitem[{{Marisaldi} {et~al.}(2015)}]{Marisaldi2015}
{Marisaldi}, M., Argan, A., Ursi, A., {et~al.} 2015, GeoRL, 42, 9481

\bibitem[{{Marisaldi} {et~al.}(2008)}]{2008A&A...490.1151M}
{Marisaldi}, M., {Labanti}, C., {Fuschino}, F., {et~al.} 2008, \aap, 490, 1151

\bibitem[{Martone} {et~al.}(2019)]{2019A&A...631A..62M}
{Martone}, R., {Guidorzi}, C., {Margutti}, R., {Nicastro}, L., {Amati}, L., {et~al.} 2019, \aap, 631, A62

%\bibitem[{{Mattox} {et~al.}(1996){Mattox}, {Bertsch}, {Chiang}, \& {et~al.}}]{1996Mattox}
%{Mattox}, J.R., {Bertsch}, D.L., {Chiang}, J., {et al.} 1996, ApJ, 461, 396.

\bibitem[{McQuinn}(2014)]{2014McQuinn}
{McQuinn}, M. 2014, ApJ, 780, L33

\bibitem[{Mereghetti}(2020)]{2020Mereghetti}
{Mereghetti}, S., {Savchenko}, V., {Ferrigno}, C., {et~al.} 2020, \apjl, 829, L29

\bibitem[{Metzger} {et al.}(2019)]{2019MNRAS.485.4091M}
{Metzger}, B.~D., {Margalit}, B.,and {Sironi}, L. 2019, MNRAS, 485, 4091

%\bibitem[{Michilli} {et~al.}(2018)]{2018Natur.533..132}
%{Michilli}, D., {Seymour}, A., {Hessels}, J.~W.~T., {et~al.}, 2018, Natur, 533, 132

\bibitem[{Nicastro} {et~al.}(2021)]{2021Univ....7...76N}
{Nicastro}, L., {Guidorzi}, C., {Palazzi}, E., {et~al.} 2021, Univ, 7, 76

%\bibitem[{Olausen} \& {Kaspi}(2014)]{2014Olausen}
%{Olausen}, S.~A., \& {Kaspi}, V.~M., 2014, ApJS, 212, 6

\bibitem[{Palmer} {et~al.}(2005)]{palmer2005}
{Palmer}, D.M., {Barthelmy}, S., {Gehrels}, N., {et~al.} 2005, Natur, 434, 1107

% Perotti!!!!!!!!!!!!!!!!!!!
\bibitem[{Perotti} {et~al.}(2006)]{2006NIMPA.556..228P}
{Perotti}, F., {Fiorini}, M., {Incorvaia}, S., {Mattaini}, E., \& {Sant'Ambrogio}, E. 2006, NIMPA, 556, 228

\bibitem[{Petroff} {et~al.}(2016)]{2016PASA33..e45}
{Petroff}, E., {Barr}, E.~D., {Jameson}, A., {et~al.} 2016, PASA, 33, e045

\bibitem[{Petroff} {et~al.}(2019)]{2019A&ARv27..4P}
{Petroff}, E., {Hessels}, J.~W.~T. \& {Lorimer}, D.~R. 2019, A\&ARv, 27, 4.

%\bibitem[{Petroff} \& {Yaron}(2020)]{2020Astron160}
%{Petroff}, E., \& {Yaron}, O., 2020, Transient Name Server AstroNote, 160, 1

% TGF HE
%\bibitem[{{Marisaldi} {et~al.}(2010){Marisaldi}, {Labanti}, {Fuschino}, \&
% {Tavani}}]{2010JGRA20106}
%{Marisaldi}, M., {Labanti}, C., {Fuschino}, F., {et~al.}, 2010, Journal of
%  Geophysical Research (Space Physics), 115, 2156

%\bibitem[{{Pilia} {et~al.}(2020a)}]{2020ATel13492}
%{Pilia}, M., {Nardi}, G., {Bernardi}, G., {et~al.}, 2020, Astronomer’s Telegram, 13492, 1

\bibitem[{Pilia} {et~al.}(2020)]{2020Pilia}
{Pilia}, M., {Burgay}, M., {Possenti}, A., {et~al.} 2020, \apjl, 896, L40, arXiv:2003.12748

\bibitem[{{Pittori}(2013)}]{2013NuPhS.239..104P}
{Pittori}, C. 2013, NuPhS, 239, 104
%{Pittori}, C. 2013, Nuclear Physics B Proceedings Supplements, 239, 104

\bibitem[{Platts} {et~al.}(2019)]{2019PhR821..1P}
{Platts}, E., {Weltman}, A., {Walters}, A., {et~al} 2019, PhR, 821, 1, http://frbtheorycat.org.

\bibitem[{Pleunis} {et~al.}(2021)]{2021ApJLPleunis}
{Pleunis}, Z., {Michilli}, D., {Bassa}, C.G., {et~al} 2021, \apjl, 911, L3

\bibitem[{Prochaska} \& {Zheng}(2019)]{2019Prochaska}
{Prochaska}, J.~X., \& {Zheng}, Y., 2019, MNRAS, 485, 648

% Konus-Wind
\bibitem[{Ridnaia}(2020)]{2020RidnaiaK-W}
{Ridnaia}, A., {Svinkin}, D., {Frederiks}, D., {et~al.} 2020, NatAs, 5, 372
%{Ridnaia}, A., {Svinkin}, D., {Frederiks}, D., {et~al.}, 2020, Nature Astronomy, 5, 372
%submitted to Natur, arXiv e-prints, arXiv:2005.11178

%\bibitem[{{Singer} \& {Price}(2016)}]{2016PhRvD..93b4013S}
%{Singer}, L.~P. \& {Price}, L.~R. 2016, Physical Review D, 93, 024013

%\bibitem[{Scholz} {et~al.}(2017)]{2017ApJ846...80}
%{Scholz}, P., {Bogdanov}, S., {Hessels}, J.~W.~T., {et~al.}, 2017, ApJ, 846, 80

\bibitem[{Scholz} (2020)]{2020ATel13681}
{Scholz}, P., 2020, ATel, 13681, 1
%{Scholz}, P., {Chime/Frb Collaboration}, {et~al.}, 2020b, Astronomer's Telegram, 13681, 1

\bibitem[{Scholz} {et~al.}(2020)]{2020ApJLScholz}
{Scholz}, P., {Cook}, A., {Cruces}, M., {et~al.} 2020, \apjl, 901, 165

\bibitem[{Spitler} {et~al.}(2016)]{2016Natur.531.202S}
{Spitler}, L.~G., {Scholz}, P., {Hessels}, J.~W.~T., {et~al.} 2016, Natur, 531, 202

% X & gamma--ray

%\bibitem[{{Spruit}(1999){Spruit}}]{1999Spruit}
%{Spruit}, H.C., 1999, A\&A, 341, L1

\bibitem[{{Tavani}(2019)}]{2019RLSFN...38}
{Tavani}, M. 2019, RLSFN, 30, 13
%, doi:10.1007/s12210-019-00841-5
%{Tavani}, M., 2019, Rend. Lincei Suppl., 30, 13, doi:10.1007/s12210-019-00841-5

\bibitem[{{Tavani} {et~al.}(2009)}]{2009A&A...502..995T}
{Tavani}, M., {Barbiellini}, G., {Argan}, A., {et~al.} 2009, \aap, 502, 995

\bibitem[{Tavani} {et~al.}(2021)]{TavaniNat2020}
{Tavani}, M., {Casentini}, C., {Ursi}, A., {et~al.} 2021, NatAs, 5, 401
%accepted, arXiv e-prints, arXiv:2005.12164

\bibitem[{{Tavani} {et~al.}(2011)}]{2011PhRvL.106a8501T}
{Tavani}, M., {Marisaldi}, M., {Labanti}, C., {et~al.} 2011, Physical Review
  Letters, 106, 018501

% AGILE GW papers
%\bibitem[{{Tavani} {et~al.}(2016)}]{2016ApJ...825L...4T}
%{Tavani}, M., {Pittori}, C., {Verrecchia}, F., {et~al.} 2016, \apjl, 825, L4

\bibitem[{Tavani} {et~al.}(2020)]{Tavani2020}
{Tavani}, M., {Verrecchia}, F., {Casentini}, C., {et~al.} 2020, \apjl, 893, L42

\bibitem[{Thompson} \& {Duncan}(1996)]{thompson1996}
{Thompson}, C. \& {Duncan}, R.~C. 1996, ApJ, 473, 322

% Paper new GW on-board config
\bibitem[{Ursi} {et~al.}(2019)] {2019ApJ...871...27U}
{Ursi}, A., {Tavani}, M., {Verrecchia}, F., {et~al.} 2019, \apj, 871, 27

%\bibitem[{{Thompson}(1994){Thompson}}]{1994Thomp}
%{Thompson}, C., 1994, MNRAS, 270, 480

%%%%%%%
%
\bibitem[{{Verrecchia} {et~al.}(2019)}]{2019RLSFN...33}
{Verrecchia}, F., {Tavani}, M.,{Bulgarelli}, A., {et~al.}, 2019, RLSFN, 30, 71
%, doi:10.1007/s12210-019-00854-0
%{Verrecchia}, F., {Tavani}, M.,{Bulgarelli}, A., {et~al.}, 2019, Rend. Lincei Suppl., 30, 71, doi:10.1007/s12210-019-00854-0

\bibitem[{{Verrecchia} {et~al.}(2017)}]{2017ApJ...850L...27V}
 {Verrecchia}, F., {Tavani}, M., {Donnarumma}, I., {et~al.} 2017, \apjl, 850, L27

%\bibitem[{{Verrecchia} {et~al.}(2018)}]{2018IAUS..338...84V}
%{Verrecchia}, F., {Tavani}, M., {Donnarumma}, I., {et~al.}, 2018, IAUS, 338, 84
%%%%%%%%%%%%%%%%%%%

%\bibitem[{{Verrecchia} {et~al.}(2017{\natexlab{a}})}]{2017ApJ...847L...20V}
% {Verrecchia}, F., {Tavani}, M., {Ursi}, A.,{et~al.} 2017{\natexlab{a}}, \apjl, 847, L20

%\bibitem[{Yao} {et~al.}(2017)]{YMW16}
%{Yao}, J.~M., {Manchester}, R.N., \& {Wang}, N., 2017, ApJ, 835, 29

%\bibitem[{{Zhang} \& {Meszaros}(2001)}]{2001Zhang}
%{Zhang}, B., and {Meszaros}, P, 2001, ApJ, 552, L35

%\end{thebibliography}

\end{thebibliography}

\end{document}